\documentclass[structabstract]{aa}  

\usepackage{graphicx}
\usepackage{txfonts}
\usepackage{natbib}
\bibpunct{(}{)}{;}{a}{}{,} % to follow the A&A style
\newcommand{\Hi}{{\sc H$\,$i} }
\newcommand{\Hii}{{\sc H$\,$ii} }

\begin{document}
   \title{ {\sc SimpleX2}: radiative transfer on an unstructured, dynamic grid}

   \author{J.-P. Paardekooper
          \inst{1}
          \and
          C.J.H. Kruip
          \inst{1}
          \and
          V. Icke
          \inst{1}
          }

   \institute{Leiden Observatory, Leiden University, Postbus 9513, 2300RA Leiden, The Netherlands\\
              \email{jppaarde@strw.leidenuniv.nl}
             }

 \date{Received 8 December 2009 / Accepted 21 March 2010}

  \abstract
   {We present an improved version of the SimpleX method for radiative transfer on an unstructured Delaunay grid. The grid samples the medium through which photons are transported in an optimal way for fast radiative transfer calculations.}
   {We study the detailed working of SimpleX in test problems and show improvements over the previous version of the method.}
   {We have applied a direction conserving transport scheme that correctly transports photons in the optically thin regime, a regime where the original SimpleX algorithm lost its accuracy. In addition, a scheme of dynamic grid updates is employed to ensure correct photon transport when the optical depth changes during a simulation. For the application to large data sets, the method is parallellised for distributed memory machines using MPI.}
   {To test the new method, we have performed standard tests for cosmological radiative transfer. We show excellent correspondence with both the analytical solution (when available) and to the results of other codes compared to the former version of SimpleX, without sacrificing the benefits of the high computational speed of the method.}
   {}


   \maketitle
%
%________________________________________________________________

\section{Introduction}
In many astrophysical applications radiative processes play an important, and occasionally dominant, role. The interaction between radiation and matter leads to possibly strong feedback effects on the medium caused by heating and cooling, radiation pressure and the change of the ionisation and excitation state of the medium. It is therefore of crucial importance to include these effects in the simulations of hydrodynamic flows. However, radiative transfer is a very complex process due to the high dimensionality of the problem. The specific intensity $I(\vec{\Omega}, \vec{x},t,\nu)$ depends on seven variables, and is impossible to solve for in an analytic way in general. With the exception of certain limiting cases where an analytical solution exists, one therefore has to rely on numerical methods to obtain the desired solution. 
\par Several kinds of methods exist for this purpose, all of which have their specific advantages and shortcomings. A relatively straightforward way of solving the radiative transfer equation is the long characteristics method (e.g.  \citet{Mihalas:1984p2300} ), where rays connect each source cell with all other cells and the transfer equation is solved along every ray. Although this method is relatively precise, it is computationally demanding, requiring $N^{2}$ interactions for $N$ cells. A solution to this unfortunate scaling is implemented in the short characteristics method (e.g. \citet{Kunasz:1988p2312} ), where only neighbouring cells are connected by rays. Not only does this make the method computationally less expensive than the long characteristics method, it is also easier to parallellise. A drawback of this method is that it is known to cause numerical diffusion in the solution. In recent years, hybrid schemes combining the benefits of both approaches have been developed \citep{Rijkhorst:2006p896,Trac:2007p1516}. Instead of direct integration along the characteristics, Monte Carlo methods use a stochastic approach by sending photon packets along the rays. The properties of the photon packets and their interaction with the medium are determined by sampling the distribution functions of the relevant processes. Moment methods solve the moments of the radiative transfer equation, which allows for a computational speed-up in certain opacity regimes. In these methods there is a trade-off between computation time and numerical diffusion in the solution, depending on what method is used to obtain the closure relation. 
\par What almost all of these methods have in common is that they use a predefined grid on which the radiative transfer calculation is done. Most often this grid is given by a hydrodynamics simulation, which is either a regular, adaptive mesh refinement (AMR) or smoothed particle hydrodynamics (SPH) grid. These grids are not optimised for radiative transfer but for hydrodynamics calculations, possibly resulting in unphysical behaviour of the numerical solution. Moreover, the computation time of almost all methods scales linearly with the number of sources, which severely limits the range of applications. Exceptions are moment methods that generally do not scale with the number of sources, but sacrifice precision by introducing numerical diffusion (e.g. \citet{Gnedin:2001p24,Cen:2002p2314}) and the method by \citet{Pawlik:2008p888}, where a source merging procedure is used to avoid linear scaling with the number of sources. For many applications, the linear scaling of the computation time with the number of sources becomes prohibitive, for example when simulating scattering processes, where effectively every cell might become a source. In the case of simulations of the epoch of reionisation, which is the topic of the second half of this paper, it is necessary to include many sources. It is therefore essential to use a method for which the computation time is independent of the number of sources. 
\par The approach to solve the radiative transfer equation taken in this paper is radically different from the methods described earlier. Instead of using a predefined grid, the {\sc SimpleX} method calculates the grid for the radiative transfer problem from the properties of the physical medium through which photons will be transported. This leads to a computationally fast method that does not scale with the number of sources, making it an ideal tool for simulations of the epoch of reionisation. A previous version of the method has been described in \citet{Ritzerveld:2006p9}, where the general idea of transport on random lattices is laid down, with a small section on the application to cosmological reionisation. The first comprehensive set of tests of the method were performed for the Radiative Transfer Comparison Project \citep{Iliev:2006p13}. In this project, the focus lay on comparing the performance of all participating codes in the test problems, making an in-depth analysis of the {\sc SimpleX} results impossible.
\par The aim of this paper is to describe the improvements of the {\sc SimpleX} method since these previous two papers. Recently, we have performed a detailed study of the error properties of the method \citep{Kruip:2009} (KPCI09), which has led to some essential improvements to the method. The two main problems in ballistic transport that were addressed in KPCI09, decollimation and deflection, are minimised both by using an alternative transport scheme in the opacity regime where these problems occur and by adapting the grid to changes in the opacity during a simulation. In addition, the algorithm has been parallellised for distributed memory. In this paper, we describe the working of the improved {\sc SimpleX} method and provide a detailed analysis of the algorithm in test problems focusing on cosmological radiative transfer. 
\par The format of this paper is as follows. In Sect.  \ref{section_simplex}, we give an overview of the {\sc SimpleX} method and a description of the new features of the method. We then describe the parallellisation strategy in Sect.  \ref{section_parallel_scheme} and present the computational scaling properties of the algorithm. In Sect.  \ref{section_cosmo_rad_trans} we focus on the specific application of the {\sc SimpleX} algorithm to the problem of cosmological radiative transfer, and describe test problems for this application in Sect.  \ref{section_test_problems}. Finally, we present a  summary in Sect.  \ref{section_conclusions}.

%__________________________________________________________________

\section{The {\sc SimpleX}  method}\label{section_simplex}
In this section, we describe the basics of the {\sc SimpleX} method and specifically the new features that were added to improve the performance in the lower opacity regimes. For the sake of clarity we repeat some essential information that was presented earlier in \citet{Ritzerveld:2006p9} and \citet{Ritzerveld:2007p1304}, which is necessary to appreciate the new features of the method. We start with a description of how the unstructured grid is created and how to optimise it for radiative transfer calculations with {\sc SimpleX}. We then proceed by describing the different ways of transporting photons on this grid, governed by the physical properties of the problem at hand.

%__________________________________________________________________

\subsection{Grid calculation}\label{section_grid}
At the basis of the {\sc SimpleX} method lies the unstructured grid on which the photons are transported. The grid adapts to the physical properties of the medium through which the photons travel in such a way that more grid points are placed in regions with a higher opacity. The result is a higher resolution in places where it's needed most, there where the optical depth is highest.

\subsubsection{Point process}
\begin{figure}
  \centering
  \includegraphics[width=9cm]{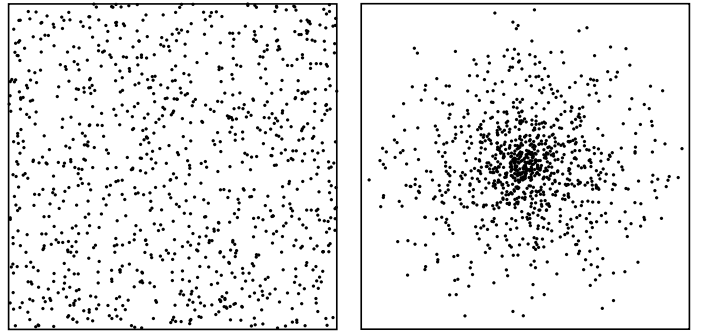}
     \caption{ Two-dimensional example of a point distribution based on a homogeneous Poisson process (\textit{left}) and based on a non-homogeneous Poisson process (\textit{right}).} 
        \label{fig_point_process}
\end{figure}

The placement of the grid points is a stochastic process based on the Poisson process, which can be defined as follows. Suppose $N(A)$ is the number of points in a non-empty subset $A$ of the volume $S \subset \mathbb{R}^{d}$, with $d$ the dimension. Then the probability to that $A$ contains $x$ points is
\begin{equation}
  \Phi = P( N(A) = x ) = \frac{n_{p} \vert A \vert e^{-n_{p} \vert A \vert x}}{x!}, \; \; \; \; \; x=0,1,2,\ldots
\end{equation}
The only parameter in this process is the point intensity $n_{p}$, which is a global constant. Every region in the volume has the same probability that points are placed there, which in our case corresponds to a constant opacity. An example of a homogeneous Poisson process is shown on the left hand side of Fig. \ref{fig_point_process}. 
\par To account for different opacity regimes inside the computational volume, we use the non-homogeneous Poisson process, defined as
\begin{equation}
  P( N(A) = x ) = \frac{n_{p}(A) \vert A \vert e^{-n_{p}(A) \vert A \vert x}}{x!}, \; \; \; \; \; x=0,1,2,\ldots
\end{equation}
where
\begin{equation}
  n_{p}(A) = \int_{A} n_{p}(\vec{x}) \mathrm{d} \vec{x}.
\end{equation}
The point intensity function $n_{p}(\vec{x})$ follows the opacity of the medium on global scale, while on local scale the point distribution retains the properties of the homogeneous Poisson distribution. An example of a point distribution based on the non-homogeneous Poisson process is shown on the right hand side of Fig. \ref{fig_point_process}. An alternative, possibly more physically intuitive, recipe for constructing the non-homogeneous Poisson process can be written as
\begin{equation}\label{point_process}
  n_p(\vec{x}) = \Phi \ast f(n(\vec{x})),
\end{equation} 
 that is, by defining the grid point distribution as a convolution of a homogeneous Poisson process and a function of the possibly inhomogeneous medium density distribution $n(\vec{ x })$. It is this recipe for the non-homogeneous Poisson process that we use to construct the {\sc SimpleX} grid. Grid points are placed by randomly sampling the correlation function $f(n(\vec{x}))$. We will discuss the exact shape of the correlation function in Sect. \ref{section_realistic_sampling}.

\subsubsection{The Delaunay triangulation}
\begin{figure*}
  \centering
  \includegraphics[width=\textwidth]{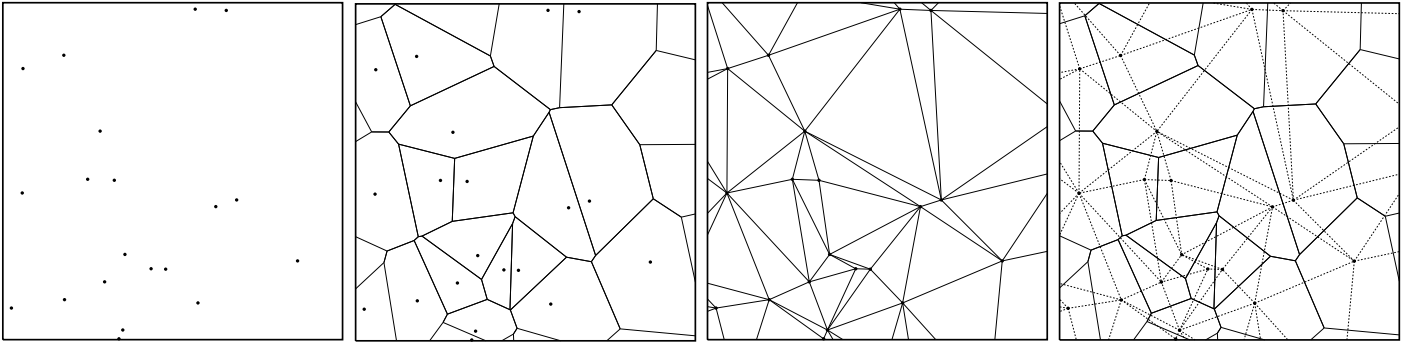}
     \caption{ Two-dimensional example of a random point distribution, the Voronoi tessellation around the points, the corresponding Delaunay triangulation connecting the points and the combination of both. }
        \label{Voronoi}
\end{figure*}

The grid points thus created form the nuclei of  a Voronoi tessellation. Given a set of nuclei ${x_i}$, the Voronoi tessellation \citep{Dirichlet:1850p2615,Voronoi:1908p2612} is defined as $V= \{C_i\}$, in which
\begin{equation}\label{voronoi_tessellation}
  C_i = \left\{ y \in \mathbb{R}^{d} : \|  x_i - y \| \le \| x_j - y \|  \, \, \forall \, x_j \ne x_i  \right\}.
\end{equation}
In other words, this means that every point inside a Voronoi cell is closer to the nucleus of that cell than to any other nucleus. By joining all nuclei that have a common facet (an edge in 2D, a wall in 3D), we create the Delaunay triangulation \citep{Delaunay:1934p2614}. Thus, every nucleus is connected to its closest neighbours. A 2D example of a Voronoi tessellation and the corresponding Delaunay triangulation is shown in Fig. \ref{Voronoi}.
\par The Delaunay triangulation consists of simplices that fill the entire domain. A simplex is the generalisation of a triangle in $\mathbb{R}^{d}$, so a triangle in $\mathbb{R}^{2}$ and a tetrahedron in $\mathbb{R}^{3}$. In a valid Delaunay triangulation, all simplices obey the empty circumsphere criterion. The circumsphere of a simplex is the unique sphere that passes through each of the vertices that make up the simplex. In a valid Delaunay triangulation, no vertex exists inside this circumsphere.
\par For Voronoi tesselations and Delaunay triangulations that are constructed from a point process based on a homogenous Poisson process, so-called Poisson Delaunay triangulations, it is possible to derive some general properties relevant for our radiative transfer method. These results were mainly derived by \citet{Miles1970,Miles1974} and \citet{Moller1989}. Two important properties for our purposes are the average number of neighbours of a vertex and the average distance between two connected vertices. The expectation value for the number of neighbours of a typical vertex in $\mathbb{R}^{2}$ and $\mathbb{R}^{3}$ is 
\begin{equation} 
  \mathrm{E}_{\mathrm{2D}}(E) = 6
\end{equation}
and 
\begin{equation} 
  \mathrm{E}_{\mathrm{3D}}(E) = \frac{48 \pi^2}{35} +2 \approx 15.54.
\end{equation}
The expectation value for the distance between two connected vertices in $\mathbb{R}^{2}$ and $\mathbb{R}^{3}$ is 
\begin{equation} \label{edge_length_2D}
  \mathrm{E}_{\mathrm{2D}}(L) = \frac{32}{9\pi} n_{p}^{-1/2} \approx 1.132 n_{p}^{-1/2}
\end{equation}
and
\begin{equation} \label{edge_length_3D}
  \mathrm{E}_{\mathrm{3D}}(L) = \frac{1715}{2304} \left(\frac{3}{4}\right)^{1/3} \pi^{-1/3} n_{p}^{-1/3} \approx 1.237 n_{p}^{-1/3}.
\end{equation}
Note that these values are only exact for Delaunay triangulations constructed from a homogeneous Poisson process, while in {\sc SimpleX} we use the non-homogeneous Poisson process to place the grid points. Except for regions in the domain with strong gradients in the point density, on local scale the point distribution resembles a homogeneous point distribution quite well. Therefore the properties of the Poisson Delaunay triangulation give a good qualitative idea of the properties of the grid on which we perform our radiative transfer calculations.

\par {\sc SimpleX} is set up in such a way that once the point distribution is created according to Eq. (\ref{point_process}), the Delaunay triangulation is calculated by an external software package. It is therefore possible to use any package that suits the application at hand. For all simulations presented in this paper, the Delaunay triangulation is calculated using the QHull package\footnote{www.qhull.org}. This is a software package written in C that is able to calculate the Delaunay triangulation, the surfaces and the volumes of the simplices in up to 8 dimensions. QHull is based on the Quickhull algorithm \citep{Barber:1995p2617}, using the convex hull property of the Delaunay triangulation. QHull has the advantages that it computes the Delaunay triangulation in optimal time $\mathcal{O} \, (N \log N)$, it is very stable against floating point round off errors in case points lie very close to each other and it is easy to implement as modular plugin routine. One of the drawbacks of QHull is that it triangulates the entire point set in one call, so it's impossible to add or delete points after the triangulation has been computed. This results in extra computational overhead in the grid dynamics scheme presented in Sect. \ref{section_grid_dynamics}. However, the computation time of the triangulation is small compared to the computation time of the radiative transfer (see also Fig. \ref{figure_scaling_1proc}), so in the present case the extra computational overhead is acceptable.

\subsubsection{The correlation function}\label{section_realistic_sampling} 
In the previous discussion we have not specified the exact shape of the correlation function $f(n(\vec{x}))$ with which we sample the density distribution of the medium. In order for the grid to adapt to the properties of the medium, the correlation function should be a monotonically increasing function in $n(\vec{x})$. Thus, the distance between two connected vertices will be smaller in regions with high density. From basic transfer theory, we know that the local mean free path in a medium relates to the local medium density in the following way:
\begin{equation}\label{mean_free_path}
  \lambda(\vec{x}) = \frac{1}{n(\vec{x} )\sigma},
\end{equation}
where $\sigma$ is the total cross section, $\sigma = \sum_{i} \sigma_j$, consisting of different interaction cross sections $\sigma_j$. If we compare this to the expectation value of the Delaunay edge length in Eq. (\ref{edge_length_2D}) and Eq. (\ref{edge_length_3D}) it follows that if we choose the correlation function $f(n(\vec{x}))$ to sample the $d$-th power of the local medium density, e.g. $f(x) \propto x^d$, the local mean free path scales linearly with the expectation value of the Delaunay edge length via a constant $c$:
 \begin{equation}\label{delaunay_vs_mfp}
 \langle L_D \rangle (\vec{x}) = c \lambda(\vec{x}).
 \end{equation}
Equation (\ref{delaunay_vs_mfp}) is a global relation with a global constant $c$, given by
\begin{equation}\label{constant_c}
  c = \xi( d, D, N ) \, \sigma.
\end{equation}
Here, $D$ is the size of the computational domain and $N$ the number of vertices.
\par This sampling recipe works very well in physical media where the density fluctuations are small. However, if the density fluctuations are significant, sampling the density to the $d$-th power will favour the high density regions, resulting in a possibly severe undersampling of the low density regions. This undersampling can have serious consequences for the radiative transfer calculation, for instance by causing preferential directions that lead radiation around low density regions. See Sect. \ref{section_test4_undersampling} for an example of this effect. Moreover, KPCI09 showed that large gradients in the grid point distribution lead to systematic errors when transporting photons on this grid. 
\par We therefore need a sampling recipe that retains the advantages of the adaptive grid by keeping the dynamic range as large as possible, while maximising the minimum resolution of the grid. This is achieved by defining a reference density $n_{0}( \vec{x} )$ above which the density is no longer sampled to the $d$-th power but a different power $\alpha$. Both $n_{0}( \vec{x} )$ and $\alpha$ depend on the properties of the medium that needs to be sampled. The two sampling recipes are smoothly joined by taking the harmonic mean, resulting in the following sampling function:
\begin{equation}\label{realistic_sampling}
  f(n(\vec{x})) = \left( \left( \frac{n(\vec{x})}{n_{0}(\vec{x})} \right)^{-d}  + \left( \frac{n(\vec{x})}{n_{0}(\vec{x})} \right)^{-\alpha} \right)^{-1}.
\end{equation}
This sampling recipe favours low density regions by sampling those with a higher ($d$-th) power. 
\par
\begin{figure}
  \centering
  \includegraphics[width=9cm]{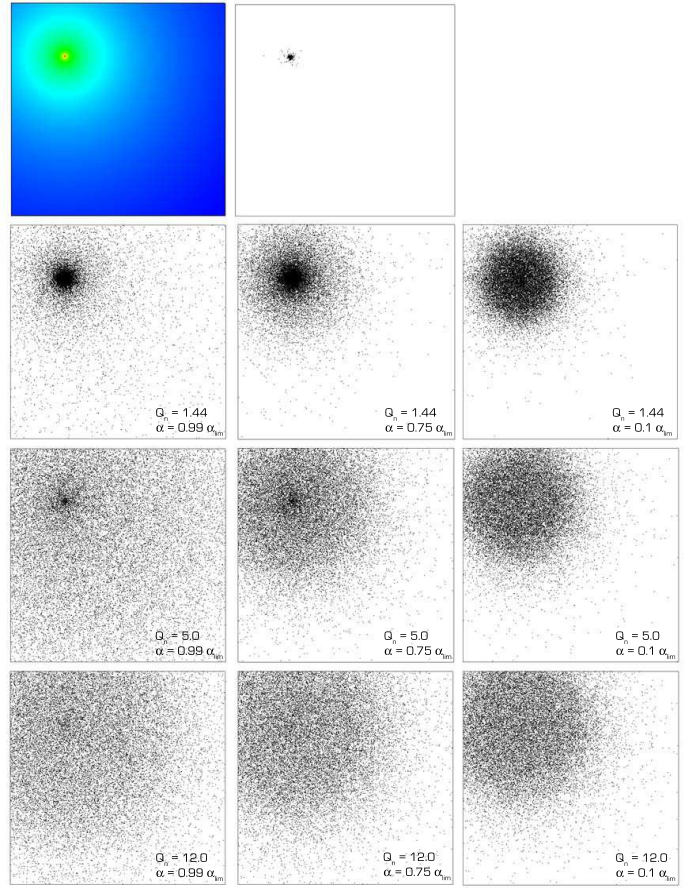}
     \caption{ Example of the influence of the sampling parameters $Q_{n}$, $\alpha$ and $n_{0}(\vec{x})$ on the sampling of a 2D density field. In all figures the number of grid points is 20000, sampling parameters are plotted such that $Q_{n}$ increases from top to bottom and $\alpha$ decreases from left to right. The values for $\alpha$ are taken relative to the limiting value for which $n_{0}(\vec{x})=0$. From left to right $\alpha$ values are $0.99 \alpha_{lim}$, $0.75 \alpha_{lim}$ and $0.1 \alpha_{lim}$. \textit{First row:} Density field with a single density peak and a $1/r^{2}$ profile with $Q_{y} = 1.44$ and the sampling according to the square of the density. Clearly this sampling recipe leads to a bias towards the high density peak. \textit{Second row:} sampling with $Q_{n}=1.44$ and $\alpha_{lim} = 1.0$. \textit{Third row:} sampling with $Q_{n}=5.0$ and $\alpha_{lim} = 0.29$. \textit{Fourth row:} sampling with $Q_{n}=12.0$ and $\alpha_{lim}=0.12$. Increasing $Q_{n}$ leads to smaller gradients in the point distribution, eventually leading to a homogeneous point distribution. Decreasing $\alpha$ implies increasing $n_{0}(\vec{x})$. This leads to a flatter point distribution close to the density peak but also to undersampling in the low density regions.}  
        \label{fig_sampling_params}
\end{figure}

 Following KPCI09 we can define a sampling parameter $Q_{n}$ that is a measure of the gradients in the point distribution:
\begin{equation}
  Q_{n} = \frac{n_{p}(\vec{x})}{\nabla n_{p}(\vec{x})}.
\end{equation}
Similarly, we can define a measure of the gradients in the density distribution:
\begin{equation}
  Q_{y} = \frac{n(\vec{x})}{\nabla n(\vec{x})}.
\end{equation}
Using Eq. (\ref{realistic_sampling}) we can find the relationship between $Q_{n}$ and $Q_{y}$:
\begin{equation}
 Q_{n} = Q_{y} \frac{ y^{-\alpha} + y^{-d}}{ y \, (\alpha y^{-\alpha - 1} +  d \, y^{-d - 1} )}, 
\end{equation}
where we have defined $y = n(\vec{x})/n_{0}(\vec{x})$. A choice for $Q_{n}$ sets the maximum gradient in the point distribution and thus the dynamic range in the simulation. The higher $Q_{n}$, the smaller the gradients in the point distribution.  An example of a density field and point distribution with different values for $Q_{n}$ is shown in Fig. \ref{fig_sampling_params}. 
This figure shows that increasing $Q_{n}$ indeed results in a smoother point distribution, with more points placed in the low density regions. A high $Q_{n}$ therefore makes it less likely that errors due to undersampling in low density regions occur. However, increasing $Q_{n}$ also decreases the dynamic range in the simulation, resulting in a possible undersampling of high density peaks for high $Q_{n}$ values. For optimal sampling one should therefore choose the lowest $Q_{n}$ value for which numerical errors due to undersampling in low density regions are within a predefined tolerance set by the requirements of the simulation. For a more elaborate discussion of the $Q_{n}$ parameter, corresponding numerical artefacts and their solutions we refer the reader to KPCI09.
\par Since $Q_{y}$ is a fixed property of the medium density, a certain value of $Q_{n}$ only exists for specific combinations of $n_{0}( \vec{x} )$ and $\alpha$. For every value of $Q_{n}$ there is a maximum $\alpha$ at which $n_{0}(\vec{x})$ goes to zero. The influence of different $\alpha$ values on the point distribution for fixed $Q_{n}$ is shown in Fig. \ref{fig_sampling_params}. A lower value of $\alpha$ implies a higher value for $n_{0}(\vec{x})$. The result is that the high density peak is less pronounced, due to the lower value of $\alpha$ while at the same time less grid points are present in the lowest density regions, due to the higher value of $n_{0}(\vec{x})$. The reason for this is that in this example there is a gradient in the medium density everywhere, so a higher $n_{0}(\vec{x})$ results in an emphasis on the regions where the density is close to this reference value. Therefore, the lowest density regions receive less points. It is therefore crucial to choose an $\alpha$ value that ensures that no strong density gradients exists at $n(\vec{x}) < n_{0}(\vec{x})$. 
\par By choosing the sampling parameter $Q_{n}$ and the reference density $n_{0}( \vec{x} )$ we can create a grid where the dynamic range is maximal without causing numerical artefacts due to undersampling of the low density regions or large gradients in the point distribution. One has to be careful that by sampling different opacity regimes in a different way, the interaction coefficient of Eq. (\ref{constant_c}) is no longer a global constant, but we can take care of this by the way the interaction at each grid point is accounted for. This will be discussed in Sect.  \ref{section_radiation_transport}.

% __________________________________________________________________

\subsubsection{A dynamic grid}\label{section_grid_dynamics}
In the previous section we described how the {\sc SimpleX} grid is created according to the properties of the medium. In this discussion, it was assumed that the medium is static and does not change during the radiative transfer calculation. However, in reality the properties of the medium change continuously under influence of, for example, gravity and radiation. For this reason, the {\sc SimpleX} grid should be updated every time step in case of full radiation hydrodynamics simulations. In this paper we do not consider the application of {\sc SimpleX} to radiation hydrodynamics but instead focus on post-processing static density fields. We will discuss the application of {\sc SimpleX} to radiation hydrodynamics simulations in future work.
\par Even though the gas density is assumed to be static, the properties of the medium might still change during a radiative transfer calculation. For example, photo-ionisation lowers the optical depth for ionising radiation. Changes in the optical depth lead to a deviation from the recipe for grid point distribution of Eq. (\ref{realistic_sampling}). In other words, the grid is no longer an optimal representation of the physical properties relevant for the radiative transfer calculation. KPCI09 showed that this might have serious consequences for the transport of photons through regions where the optical depth between grid points is much lower than unity. In Sect. \ref{section_direction_conserving_transport} we describe a new transport scheme that minimises errors in the optically thin regime. This section describes a different solution for transport through regions where the optical depth has severely changed, the removal of grid points.
\par One of the advantages of the adaptive grid that {\sc SimpleX} uses is that resolution is put where it is needed most, in the regions with highest optical depth. If during a radiative transfer calculation the optical depth changes, we are effectively wasting computational resources in the regions with high resolution where the opacity has decreased. The high resolution is no longer necessary, since no interesting radiative transfer effects that need to be resolved at high resolution are taking place. The superfluous grid points only add to the computation time. We therefore remove unnecessary grid points from the regions where the optical depth has significantly decreased.
\par Another reason for the removal of superfluous grid points is that the transport of photons between grid points with an optical depth lower than unity is prone to numerical errors. KPCI09 showed that photons that travel through these regions are subject to decollimation and deflection. These effects are caused by the grid that is no longer an optimal representation of the properties of the medium. A straightforward solution for these problems is updating the grid in such way that the physical properties of the medium remain correctly accounted for. By ensuring that the optical depth between grid points is always close to unity, the grid remains optimally suited for the radiative transfer calculation.
\par In some cases the removal of grid points according to this scheme leads to regions devoid of grid points. An example is the photo-ionisation of a cloud of neutral hydrogen gas, which causes the optical depth for ionising photons to drop so dramatically that almost no grid points should be placed if the optical depth between grid points has to be of order unity. This extreme example leads to different errors in the solution than the ones previously described. For example, recombination rates will be incorrect as the density in the cloud is no longer resolved by any grid point. We circumvent this by imposing a minimum resolution below which no grid points are removed. This ensures that in every part of the simulation relevant structures are resolved and our requirements for accuracy are met. The effect of grid point removal and the optimal value for the minimum resolution in realistic applications will be further explored in Sects. \ref{section_test1_sc} and \ref{section_test4_dynamics}.
\par The consequence of preventing undersampling errors is that there will remain grid points in the simulation domain between which the optical depth is lower than unity. Numerical errors in the transport of photons between these grid points are prevented by using the transport scheme described in Sect. \ref{section_direction_conserving_transport}. As we will show in that section, this transport scheme is more expensive both in computation time as in memory usage compared to the other transport schemes. For optimal computation time and memory usage it is therefore important to keep the number of superfluous grid points in regions with low opacity to a minimum.

%__________________________________________________________________

\subsection{Radiation Transport}\label{section_radiation_transport}
We have shown how the unstructured grid is created on which the radiative transfer calculation will be performed. In this section we will show how we can employ the unique properties of this grid to efficiently and accurately solve the radiative transfer equation.
\par During a radiative transfer calculation photons are transported from grid point to grid point along the edges of the Delaunay triangulation. At every grid point an interaction takes place, with the interaction coefficient given by Eq. (\ref{constant_c}). According to the solution of the 1-dimensional radiative transfer equation the number of photons that interacts at this grid point is
\begin{equation}\label{photon_abs}
N_{act} = N_{in}(1-e^{-c}),
\end{equation}
where $N_{in}$ is the number of incoming photons. The number of photons that is not interacting at the grid point is
\begin{equation}\label{photon_out}
N_{out} = N_{in}e^{-c},
\end{equation}
these photons should continue along their original path. 
\par In this photon transport scheme there is no difference between a grid point that is a source and a grid point that is not. A source is simply defined as a grid point that sends photons to all its neighbours, which has no influence on the number of computations involved. Thus, the {\sc SimpleX} method does not scale with the number of sources. 
\par After interaction at a grid point there are three ways of photon transport, depending on the opacity regime and the physical process at hand. 

\subsubsection{Scattering processes} \label{section_diffuse_transport}
\begin{figure*}
  \centering
  \includegraphics[width=\textwidth]{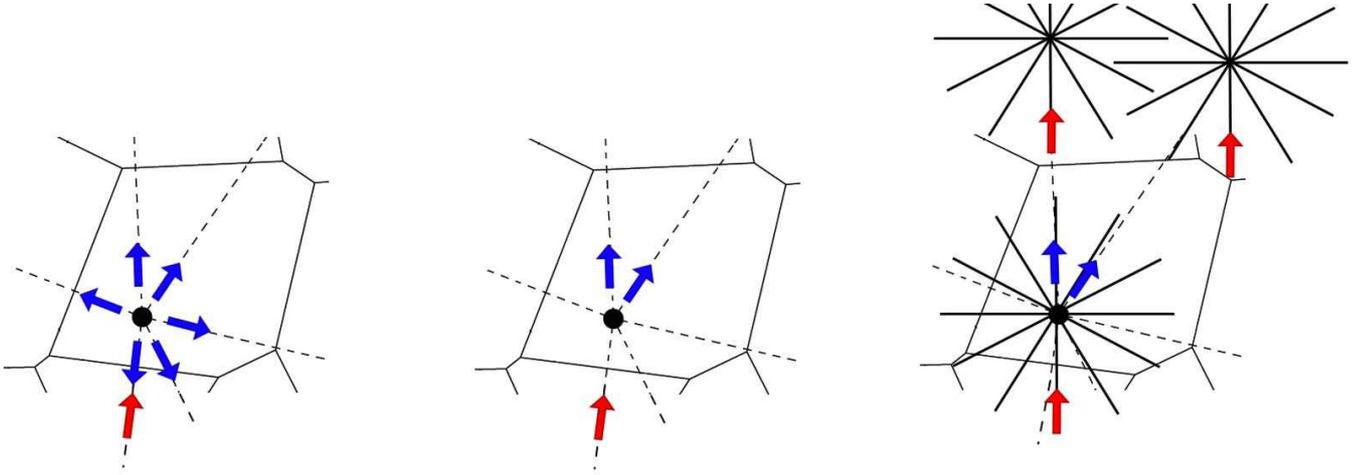}
     \caption{Two-dimensional examples of the three modes of transport with which photons are transported from one grid point to another. Red arrows indicate incoming photons, blue arrows outgoing photons. Transport in case of a scattering process is shown on the left hand side. Incoming radiation loses all memory of the initial direction and is sent to all neighbours of the vertex. Ballistic transport is shown in the centre plot. The photons are redistributed to the 2 most straightforward neighbours of the vertex, with respect to the Delaunay edge of the incoming photons. On the right hand side direction conserving transport is shown. The photons are redistributed to the 2 most straightforward neighbours with respect to the Delaunay edge that is associated with the direction bin the incoming photons are in. The outgoing photons stay in the same direction bin and have thus a memory of their original direction. }
        \label{3_types_of_transport}
\end{figure*}

In the case of an isotropic scattering process, or the absorption and re-emission of photons at a grid point, the outgoing photons have no memory of their original direction. In {\sc SimpleX} these photons are isotropically redistributed to all neighbouring vertices, as depicted on the left hand side of Fig. \ref{3_types_of_transport}. 
This type of transport does not increase the number of computations involved, {\sc SimpleX} is therefore ideally suited to simulate scattering processes.
\par For the application to ionising radiation it is straightforward to include the diffuse radiation from recombining atoms in the simulation. Hydrogen ions that capture an electron directly to the $n=1$ state emit photons capable of ionising other atoms. Most radiative transfer methods do not include this radiation but instead adopt the on-the-spot approximation (e.g. \citet{Osterbrock:1989p2350}), assuming these photons to be absorbed close to the emitting atom (see also Sect.  \ref{ionisation_proc}). Even though the validity of this approach is not well established \citep{Ritzerveld:2005p17} we will use the on-the-spot approximation for all the tests presented in this paper, in order to make a clean comparison between our results and the analytic solution or the results of other codes that all use the on-the-spot approximation. 

\subsubsection{Ballistic transport} \label{section_ballistic_transport}
The $N_{out}$ photons in Eq. \ref{photon_out} that are not interacting at a grid point should continue travelling in their original direction. However, on the unstructured grid there is no outgoing Delaunay edge in the same direction as the incoming edge. This is solved by splitting the photon packets into $p$ equal parts and dividing these packets among the $p$ most straightforward neighbours with respect to their incoming direction. The total number of photons is conserved in this process. Tests indicate that if we choose $p$ equal to the dimension of the problem $d$, the solid angle that is represented by one Delaunay edge of the emitting vertex is best preserved. In other words, a source that sends out photons in all directions will fill the entire 'sky' with radiation. However, on the unstructured grid it might be the case that one of the most straightforward neighbours lies outside a cone of 90 degrees. To prevent photons form travelling backwards, we exclude those neighbours. In Fig. \ref{3_types_of_transport} the centre plot shows an example of ballistic transport. 
\par The advantage of this transport scheme is that the most straightforward directions have to be calculated only once, at the start of the simulation. As long as the grid doesn't change these directions do not have to be recalculated. One important disadvantage of this transport scheme is that there is no memory of the original direction of the photons. At every interaction the outgoing direction was computed from the incoming direction in that step, so deflections from the original direction can add up, causing numerical diffusion to dominate after approximately 5 interactions (KPCI09). As long as the mean free path of photons is smaller than 5 Delaunay edges, this numerical diffusion has no influence on the results, since photons will be interacting with the medium before the diffusion becomes dominant. This means that during the grid calculation we have to be careful that the interaction coefficient $c$ in Eq. (\ref{constant_c}) is close to one or larger. However, this may lead to too severe constraints on the number of grid points that can be placed in optically thin regions. Therefore, a different type of transport can be employed in optically thin media.

\subsubsection{Direction conserving transport}\label{section_direction_conserving_transport}
If the interaction coefficient $c$ in Eq. (\ref{constant_c}) is smaller than one, it is no longer sufficient to determine the direction of the photons from the direction in the previous step, but a memory of the initial direction of the photon is needed. If every photon would remember its initial direction and at every interaction point the next interaction point would have to be calculated from this direction, the computation time in optically thin regions would grow unacceptably. Instead, the original direction is preserved by confining photons to solid angles corresponding to global directions in space. Unless interacting with the grid, photons stay in the same solid angle as they travel along the grid. Even though the direction of the photons is now decoupled from the grid, the photons still travel along the edges of the triangulation in the same manner as during ballistic transport. Direction conserving transport is shown on the right hand side of Fig. \ref{3_types_of_transport}.
\par
\begin{figure}
  \centering
  \includegraphics[width=8cm]{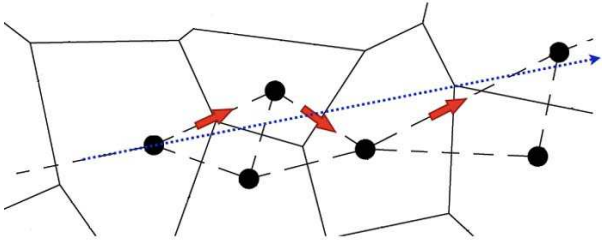}
     \caption{Example of the photon path deviating from a straight line due to the unstructured grid. Photons that are travelling in the direction of the Delaunay edge coming in from the left, should be travelling in a straight line along the dotted blue line. However, as this is impossible on the unstructured grid, photons travel along the Delaunay edges closest to their original direction. This path is depicted by the red arrows. The total path depicted by the red arrows is longer than the length of the blue line, which is corrected for by a global factor.}
        \label{photon_path}
\end{figure}

 Since photons still travel along the edges of the triangulation, the photon path deviates from an exact straight line in which the photons should be travelling, see Fig. \ref{photon_path}. 
Even though the direction of the photons is preserved, their paths are longer than physically possible. In other words, photons travel slower than the light speed on the unstructured grid. We solve this problem by applying a global correction factor to the distance between grid points, thus ensuring photons travel with the correct speed. 
\par Introducing these global directions on the unstructured grid gives rise to preferential directions, one of the problems the unstructured grid was meant to solve. By rotating the solid angles over random angles in between photon transport, the preferential directions disappear. The drawback of this procedure is that it makes direction conserving transport computationally more expensive than ballistic transport. For the latter, the most straightforward directions are calculated from the grid and thus have to be calculated only once, at the start of the simulation. For direction conserving transport it is necessary to recalculate the most straightforward directions every time the direction bins rotate. Another drawback of direction conserving transport is that the photons now have to be stored in $n$ direction bins instead of on average 16 neighbours. Typically, $n=42$ gives converged results, but as we will see in Sect. \ref{section_shadowing} this depends on the number of optically thin grid points the photons traversed. Thus, the memory requirements for direction conserving transport are higher.

\subsubsection{Combined transport}\label{section_combined_transport}
The three modes of transport described above are in general applied simultaneously during a simulation. Depending on the physical process at hand, photons are transported to all neighbours (diffuse transport), or to the $d$ most straightforward neighbours (ballistic or direction conserving transport). In regions where the optical depth is higher than or close to one, ballistic transport is used, while in the optically thin regions direction conserving transport is applied. 
\par Of the three modes of transport, direction conserving transport is computationally the most expensive (see Sect.  \ref{section_scaling} for a comparison between the computation time of ballistic and direction conserving transport). By applying this scheme only in the regions where it is necessary, the computation time is drastically reduced. As mentioned earlier, numerical diffusion starts to dominate in ballistic transport after approximately 5 steps. A first guess would therefore be to switch from ballistic to direction conserving transport when the optical depth after 5 interactions is less than one. That way, we are sure that the majority of photons does not take more than 5 steps in ballistic transport. The influence of the optical depth at which is switched in a realistic simulation is studied in more detail in Sect. \ref{test1_CT_switchTau}. Another way to reduce the computation time is by applying the grid dynamics scheme from Sect. \ref{section_grid_dynamics}. Removing superfluous grid points in the low opacity regime limits the number of vertices at which direction conserving transport is performed.
\par In combined transport we need to convert from one transport scheme to another. This is straightforward because every Delaunay edge of an optically thin vertex is associated with a solid angle in a global direction. When this vertex sends photons to an optically thick vertex the photons are transported along the Delaunay edges, so the optically thick vertex stores the photons according to the Delaunay edge associated with the solid angle. In the opposite case, when an optically thick vertex sends photons to an optically thin vertex, the photons are converted to the solid angles associated with the Delaunay edge along which the photons were sent. 

%__________________________________________________________________
\section{Parallellisation strategy}\label{section_parallel_scheme}
Even though the radiative transfer scheme presented in the previous sections is computationally efficient, in order to do large simulations involving a very high number of grid points it is necessary that the algorithm can run in parallel on distributed memory machines. This will not only reduce the computation time involved, it also reduces the amount of memory needed at each processor to store the physical properties of the grid points. The transport algorithm described in Sect.  \ref{section_radiation_transport} has the advantage that it is local: the only information needed to do a radiative transfer calculation is stored at the neighbours of the vertex. This makes the method relatively easy to parallellise. By choosing a smart domain decomposition we can minimise the number of communications involved.
 
\subsection{Domain decomposition}
The computation time of a {\sc SimpleX} calculation is independent of the number of sources, it is therefore sufficient to have a domain decomposition that assigns every processor an approximately equal number of grid points. Dividing space into equal volumes and assigning each volume to a processor is not sufficient because the number of points in each volume may differ dramatically due to the adaptive grid. We therefore use a domain decomposition based on the space-filling Hilbert curve, which is also employed in other methods without a regular grid \citep{Shirokov:2005p2410,Springel:2005p2411,Springel:2010p1514}. The Hilbert curve is a fractal that completely fills a cubic rectangular volume. A Hilbert curve is uniquely defined by its order $m$ and its dimensionality $d$, filling every cell of a $d$-dimensional cube of length $2^{m}$. The following properties of the Hilbert curve are beneficial when using it for domain decomposition:
\begin{itemize}
\item Locality: points that are close along the 1D Hilbert curve are in general also close in 3D space.
\item Compactness: a set of cells defined by a continuous section of the Hilbert curve has a small surface to volume ratio.
\item Self-similarity: the Hilbert curve can be extended to arbitrarily large size.
\end{itemize}
The first two properties minimise the number of communications between processors, while the third property ensures that we can use an arbitrarily large number of cells to determine the domain decomposition.
\par The first step in the domain decomposition is dividing the domain into $2^{md}$ equal, regular cells, where $d$ is again the dimension and $m$ the order of the Hilbert curve. We then step through the cells along the Hilbert curve, counting the number of grid points inside each cell until the number of grid points equals the total number of grid points divided by the number of processors. In this way, every processor approximately holds an equal number of grid points, thus dividing the work load evenly, while the necessary communications between processors are minimal due to the locality and compactness property of the Hilbert curve.

\subsection{Parallel radiative transfer}
The QHull algorithm that is applied to calculate the triangulation works only in serial. In the case of parallel execution of {\sc SimpleX}, every processor calculates the triangulation of the vertices that belong to that processor and vertices in a border around it belonging to neighbouring processes. This border is used to connect the triangulation between different processes. We ensure that the border is large enough by using the empty circumsphere principle. For all simplices that contain at least one vertex inside the domain of the current process, we ensure that the circumsphere of the simplex lies entirely within the boundary around the domain. Thus, we are certain that no vertex exists on another process that lies inside the circumsphere of this simplex and the triangulation is valid. After the triangulation algorithm has been applied, every process keeps only the vertices assigned to the process and a local copy of vertices assigned to other processes that are neighbour to a vertex on this process. These local copies are strictly used to send photons from one process to another, no physical interaction is taking place.  
\par The transport scheme described in Sect.  \ref{section_radiation_transport} is local, because photons are only transported from one grid point to another. We can therefore do a full radiative transfer time step without any communication between processes. During a time step, photons might be sent to a neighbour of a vertex that does not exist on the current process. After each time step, these photons are then communicated to the appropriate processes and the cycle starts anew. The local copies of vertices are only used for the transport of photons, all the physical interactions are taking place on the process that the vertex is assigned to. Hence, we are certain that physical interactions take place exactly once for every vertex during a radiative transfer time step.

\subsection{Scaling tests}\label{section_scaling}
To check how well the described parallellisation strategy works, we have conducted several scaling tests. The test consists of a simulation of a single source in a homogeneous medium, similar to the set-up of the test in Sect.  \ref{section_test1} with the difference that the simulation is stopped at 10 Myr. No grid points were removed during these tests. All the simulations were conducted using AMD Opteron 246 64Bit CPUs of 2.6 Ghz with 4 GB of memory per node. Unfortunately, we had only 8 nodes available for these tests, but it will give a general idea of the scaling. Note that even though we have used only one source for these scaling tests, the inclusion of more sources would not have influenced the timings presented here. 
\par
\begin{figure}
   \centering
   \includegraphics[width=8cm]{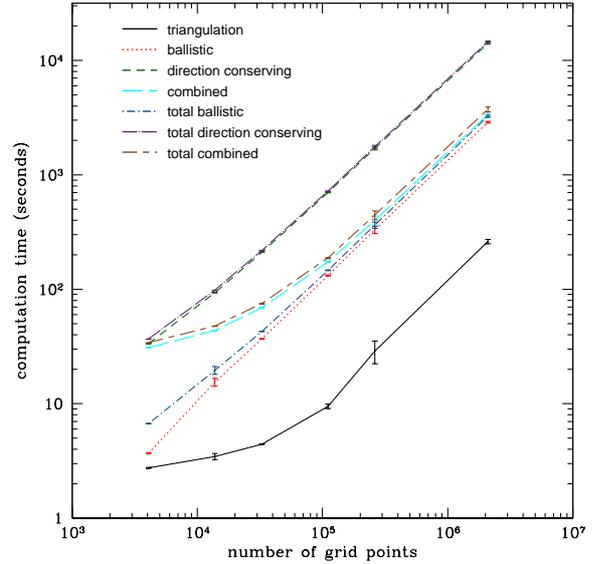}
      \caption{Simulation time as a function of the number of grid points. Shown are the computation time of the triangulation (black solid curve), ballistic radiation transport (red dotted), combined radiation transport (cyan long dashed), direction conserving transport (green short dashed), total simulation time with ballistic transport (blue dot-short dashed), total simulation time with combined transport (brown short dash-long dashed) and total simulation time with direction conserving transport (violet dot-long dashed). A doubling of the number of grid points $N$ means a doubling of the computation time for most parts of the simulation. An exception is the triangulation algorithm that scales $\mathcal{O} \, (N\log N)$ and the combined transport scheme. The computation time of the latter depends strongly on the number of optically thin grid points and thus on the physics of the simulation at hand, but will always be in between the computation time of ballistic transport and direction conserving transport. }
         \label{figure_scaling_1proc}
\end{figure}

 As a first test we used only one node with an increasing number of grid points, shown in Fig. \ref{figure_scaling_1proc}.
This figure shows that an increase of the number of grid points $N$ increases the computation time linearly for most components of the simulation. Exceptions are the triangulation algorithm, which scales $\mathcal{O} \, (N \log N)$ and the combined transport scheme. The computation time of the combined scheme will always be between the computation time of ballistic transport and that of direction conserving transport. It depends highly on the number of optically thin grid points. For the low resolution simulations, the number of optically thin grid points is relatively high and therefore the computation time is comparable to the computation time of direction conserving transport. Increasing the number of grid points decreases the relative number of optically thin grid points and therefore the computation time of combined transport comes closer to that of ballistic transport in the case of more grid points. Note that this is a feature of the set-up that we chose for the scaling test. If a larger region of the computational volume would be ionised, the computation time would be longer. On the other hand, if the ionised region were smaller, the computation would be shorter. Fig. \ref{figure_scaling_1proc} also shows that the computation time of the entire simulation is dominated by the radiation transport and not by the construction of the triangulation, since the computation time of the triangulation is one order of magnitude shorter than that of the radiation transport.
\par
\begin{figure}
   \centering
   \includegraphics[width=8cm]{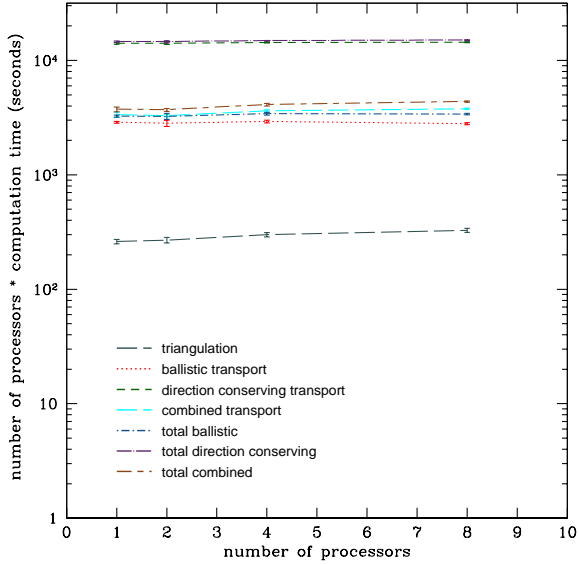}
      \caption{Simulation time as a function of the number of processors for a constant number of grid points. The number of grid points is $128^{3}=2097152$. Shown are the computation time of the triangulation (slate dashed curve), ballistic radiation transport (red dotted), combined radiation transport (cyan long dashed), direction conserving transport (green short dashed), total simulation time with ballistic transport (blue dot-short dahed), total simulation time with combined transport (brown short dash-long dashed) and total simulation time with direction conserving transport (violet dot-long dashed). Most components of the simulation scale very close to linear as the number of processors increases. An exception is the triangulation algorithm, due to the fact that every processor needs to triangulate extra points that are in the boundary between processors.}
         \label{figure_scaling_strong}
\end{figure}

 Fig. \ref{figure_scaling_strong} shows the strong scaling properties of the {\sc SimpleX} algorithm. 
We simulate the same physical problem as the previous test, but this time the number of grid points is held constant at $128^{3} = 2097152$. By increasing the number of processors we can analyse the extra work that needs to be done when more processors are employed. In the ideal case no extra work would have to be done at all, this would result in the black solid curve shown in the figure. The only component of the simulation that does not follow the linear scaling very well is the triangulation, which can be easily understood by the way the triangulation is constructed on multiple processors. Every processor has to calculate the triangulation of the grid points assigned to the processor and an additional number of grid points in a boundary around this domain. Increasing the number of processors thus means effectively increasing the number of grid points that needs to be triangulated and therefore the scaling is not as favourable as one might hope. However, the computation time of the triangulation remains an order of magnitude smaller than the radiative transfer components that do scale almost linearly, so this presents no serious issue.
\par
\begin{figure}
   \centering
   \includegraphics[width=8cm]{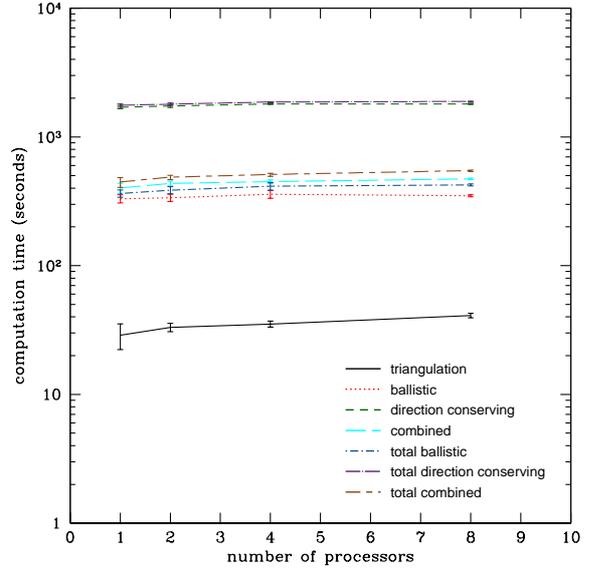}
      \caption{Simulation time as a function of the number of processors for a constant number of grid points per processor. The number of grid points at each processor is $64^{3}=262144$. Shown are the computation time of the triangulation (black solid curve), ballistic radiation transport (red dotted), combined radiation transport (cyan long dashed), direction conserving transport (green short dashed), total simulation time with ballistic transport (blue dot-short dahed), total simulation time with combined transport (brown short dash-long dashed) and total simulation time with direction conserving transport (violet dot-long dashed). The computation time of the radiation transport components shows a marginal increase as the problem size gets bigger due an increase in the number of communications involved. The computation time of the triangulation increases more for the same reason as before, that more boundary points have to be triangulated when the number of processors increases.}
         \label{figure_scaling_weak}
\end{figure}

 Finally, Fig. \ref{figure_scaling_weak} shows the weak scaling of the algorithm.
With the same problem set-up as before, we now increase the number of processors while keeping the number of grid points per processor constant. Ideally, the amount of work per processor would stay the same and the computation time would remain constant. Within a few percent the computation time for the radiation transport components remains constant, it only increases marginally due to an increase in the amount of communication necessary. As with the strong scaling, the triangulation algorithm needs to triangulate more boundary points as the number of processors increases and therefore the computation time increases with number of processors. However, the number of boundary points is small compared to the total number of grid points that needs to be triangulated, therefore the curve of the computation time flattens with an increase of the number of processors.

%__________________________________________________________________

\section{Radiative transfer of ionising radiation}\label{section_cosmo_rad_trans}
Having laid down the basics of the transport mechanism of {\sc SimpleX}, in this section we will describe the application to ionising radiation. Generally, the input for the radiative transfer calculation will be given by a hydrodynamics simulation that can be in the form of a regular, AMR or SPH grid, from which the {\sc SimpleX} grid is calculated according to the recipe described in Sect. \ref{section_grid}. Photons are then transported on this grid as described in Sect. \ref{section_radiation_transport}. Our aim is to apply {\sc SimpleX} to cosmological applications, we therefore start this section with the cosmological radiative transfer equation. However, the method can be used for all applications that require the transport of ionising photons.

\subsection{Cosmological radiative transfer equation}\label{section_transfer_equation}
The cosmological radiative transfer equation reads (e.g. \citet{Gnedin:1997p2439,Norman:1998p2308})
\begin{equation}\label{cosmological_transfer_equation}
  \frac{1}{c} \frac{\partial I_{\nu} }{ \partial t} + \frac{ \vec{\hat{n}} \cdot \nabla I_{\nu} }{ \bar{a} } - \frac{H(t)}{c}\left(  \nu \frac{ \mathrm{d} I_{\nu}}{ \mathrm{d} \nu } - 3 I_{\nu} \right) = j_{\nu} - \alpha_{\nu}I_{\nu},
\end{equation}
where $I_{\nu} = I(\vec{\hat{n}}, \vec{x},t,\nu)$ is the specific intensity with frequency $\nu$ along the unit propagation direction vector $\hat{n}$ at position $\vec{x}$ and time $t$, $j_{\nu}(\vec{x}, \vec{\hat{n}})$ is the emission coefficient, and $\alpha_{\nu}(\vec{x}, \vec{\hat{n}})$ is the extinction coefficient. The latter includes the scattering coefficient $\alpha_{\nu}^{scat}(\vec{x}, \vec{\hat{n}})$ and the pure absorption coefficient $\alpha_{\nu}^{abs}(\vec{x}, \vec{\hat{n}})$. Furthermore, $H(t) \equiv \dot{a}/a$ is the time-dependent Hubble parameter, and $\bar{a} = \frac{1+z_{em}}{1 + z}$ is the ratio of cosmic scale factors between photon emission and present time $t$. If the mean free path of photons is much smaller than the horizon, the classical transfer equation is a valid approximation:
\begin{equation}\label{transfer_equation}
  \frac{1}{c} \frac{\partial I_{\nu} }{ \partial t} + \vec{\hat{n}} \cdot \nabla I_{\nu} = j_{\nu}(\vec{x}, \vec{\hat{n}}) - \alpha_{\nu}(\vec{x}, \vec{\hat{n}})I_{\nu}.
\end{equation}
This approximation holds fairly well during the beginning of reionisation. However, care must be taken when ionised bubbles start to overlap and the photon mean free path increases dramatically. In this case the expansion of the Universe does have to be taken into account. 
\par One more simplification can be made if $j_{\nu}$ and $\alpha_{\nu}$ change on time scales larger than the light crossing time in the simulation volume. In this case the time-dependence can be dropped:
\begin{equation}\label{transfer_equation_time_independent}
 \vec{\Omega} \cdot \nabla I_{\nu} = j_{\nu}(\vec{x}, \vec{\hat{n}}) - \alpha_{\nu}(\vec{x}, \vec{\hat{n}})I_{\nu}.
\end{equation}
It is this equation that {\sc SimpleX} solves. In most astrophysical fluid flows this approximation holds fairly well as long as time scales over which the radiative transfer equation is solved are sufficiently short. However, one must take care that this equation implicitly assumes an infinite speed of light, which might result in unphysical behaviour. For example, as was pointed out by \citet{Abel:1999p992} close to ionising sources an ionisation front might travel faster than the speed of light. In the tests presented in this paper, we have checked the validity of this approximation.

\subsection{Ionisation processes}\label{ionisation_proc}
Photons with frequencies above the Lyman limit $\nu_{0}$ can be absorbed by neutral hydrogen atoms. The number of photoionisations per unit time per hydrogen atom is \citep{Osterbrock:1989p2350}:
\begin{equation}\label{photoionisation_rate}
  \Gamma_{\mathrm{ H}} = \int_{\nu_{0}}^{\infty} \frac{4 \pi J_{\nu}}{h \nu} \sigma_{\mathrm{H}}(\nu) \mathrm{d} \nu,
\end{equation}
where $J_{\nu}$ is the mean intensity and $\sigma_{\mathrm{H}}(\nu)$ is the ionisation cross section for H by photons with energy $h \nu > h \nu_{0}$. This cross section can be approximated by
\begin{equation}\label{cross_section}
   \sigma_{\mathrm{H}}(\nu) = A_{0} \left( \frac{\nu_{0}}{\nu} \right)^{3} \, \, \, \forall \, \nu \ge \nu_{0},
\end{equation}
with reference value $A_{0} = \sigma_{\mathrm{H}}(\nu_{0}) = 6.3 \cdot 10^{-18} \, \, \mathrm{cm}^{2}$. The inverse process is the recombination of electrons with the ions. The number of recombinations per hydrogen atom per unit time is:
\begin{equation}\label{recombination_rate}
  R_{\mathrm H} = n_{e} \alpha_{\mathrm{H}}(T),
\end{equation}
where $\alpha_{\mathrm{H}}(T)$ is the recombination coefficient of hydrogen atoms and $n_{e}$ the electron density. The evolution of the number density of ionised atoms at a certain point in space can then be written as
\begin{equation}\label{rate_eq}
\frac{\mbox{d}n_{ \mathrm{  H\,II} }}{\mbox{d}t} = n_{\mathrm{ H\,I}} \Gamma_{\mathrm{ H}}  - n_{\mathrm{ H\,II}}R_{\mathrm{H}},
\end{equation} 
where $n_{ \mathrm{  H\,II}}$,  $n_{ \mathrm{  H\,I}}$ and $n_{ \mathrm{  H}}$ are respectively the number density of ionised hydrogen atoms, neutral hydrogen atoms and the total number of atoms. In this equation we have neglected collisional ionisations, but they can be included trivially by adding an extra term to the equation. Relevant time scales for these processes are the photoionisation time scale $t_{ion} \equiv 1/\Gamma$ and the recombination time scale $t_{rec} \equiv 1/R$.
\par When a hydrogen ion captures an electron directly to the ground level, radiation is emitted with a frequency above the Lyman limit. In almost all radiative transfer codes, it is assumed that this radiation is absorbed close to the emitting atom, the so-called on-the-spot approximation. One can therefore ignore recombinations to the ground level, as they are cancelled out by the emitted radiation, and use the corresponding recombination coefficient $\alpha_{B}(T)$. However, \citet{Ritzerveld:2005p17} showed that, depending on the density distribution, if the source produces radiation just above the Lyman limit, this approximation is not valid. Most radiative transfer codes are incapable of including this diffuse recombination radiation since effectively every grid cell that is ionised becomes a source. In the {\sc SimpleX} algorithm it is straightforward to include the diffuse recombination radiation self-consistently. However, the analytic solutions and the results from other codes with which we will be comparing the {\sc SimpleX} results in the next section all rely on the on-the-spot approximation. Therefore, we will use the on-the-spot approximation for all tests presented in this paper.

\subsection{Assigning sources}
On the {\sc SimpleX} grid, a source is defined as a grid point that sends photons to all of its neighbours. For all the tests presented in this paper, we use a single frequency for the photons. The luminosity of the source is obtained by integrating the source spectrum $S_{\nu}$ divided by the energy of each photon over frequencies higher than the Lyman limit $\nu_{0}$:
\begin{equation}\label{source_luminosity}
  L_{ion} = \int_{\nu_{0}}^{\infty} \frac{S_{\nu}}{h \nu} \mathrm{d} \nu.
\end{equation}
By integrating the source spectrum over frequency, we neglect the influence of the spectrum on the interaction of the photons with the medium. We compensate for this by using a mean opacity representation (see \cite{Mihalas:1984p2300}):
\begin{equation}\label{mean_opacity}
  \bar{\sigma} = L_{ion}^{-1} \int_{0}^{\infty} \mathrm{d} \nu \frac{S_{\nu}}{h \nu} \sigma_{\nu}.
\end{equation}
The number of photons that a source sends out at every step is determined by the source luminosity and the radiative transfer time step $\Delta t_{rt}$.

\subsection{Interaction}
Photons that were sent out by a source travel in one radiative transfer time step from grid point to grid point, where an interaction with the medium takes place. The photons travel a distance $\Delta L$ between grid points, and thus encounter an optical depth
\begin{equation}\label{optical_depth}
  \Delta \tau = n_{\mathrm{ H\,I}} \bar{\sigma} \Delta L = (1-x)n_{\mathrm{ H}} \Delta L,
\end{equation}
in which $x$ is the ionised fraction inside the Voronoi cell through which the photon travels. Except for the ionised fraction, all these quantities have been calculated during the creation of the grid. Eq. (\ref{optical_depth}) is equivalent to the interaction coefficient in Eq. (\ref{constant_c}). Thus, if the incoming number of ionising photons is $N_{in}$, the number of ionising photons that is absorbed at this grid point is
\begin{equation}\label{intensity_absorbed}
N_{abs} = N_{in}(1-e^{-\Delta \tau}),
\end{equation}
and the number of ionising photons that is propagating onwards is
\begin{equation}\label{intensity_out}
N_{out} = N_{in}e^{-\Delta \tau}.
\end{equation}
The number of photons that are absorbed ionise the medium, thereby changing the local ionised fraction. As the medium gets ionised, the optical depth at this grid point changes, which means we should either use the direction preserving scheme at this grid point, or remove grid points to preserve the relation between optical depth and Delaunay line length, as described in Sect. \ref{section_radiation_transport} and Sect. \ref{section_grid_dynamics}. 

\subsection{Solving the photoionisation rate equation}
Having established the number of photons that is available for ionising the cell, it is straightforward to convert this to a photoionisation rate and solve Eq. (\ref{rate_eq}). However, care must be taken since by doing this we implicitly assume that the neutral density stays constant during a radiative transfer time step. This is only true for $\Delta t_{\mathrm{rt}} \ll t_{\mathrm{ion}}$ and $\Delta t_{\mathrm{rt}} \ll t_{\mathrm{rec}}$. We therefore adopt the scheme described in \citet{Pawlik:2008p888} and subcycle the rate equation on time steps $\Delta t_{\mathrm{chem}} \le \Delta t_{\mathrm{rt}}$ assuming that the ionising flux is constant during a radiative transfer time step. This ensures photon conservation even if the radiative transfer time steps are large. The rate equation at time $t_{\mathrm{chem}} \in \left( t_{\mathrm{rt}}, t_{\mathrm{rt}} + \Delta t_{\mathrm{rt}}  \right)$ is then 
\begin{equation}\label{rate_eq_subcycle}
    \mbox{d}n_{ \mathrm{  H\,II} }^{(t_{\mathrm{chem}})} = n_{\mathrm{ H\,I}}^{(t_{\mathrm{chem}})} \Gamma_{\mathrm{ H}}^{(t_{\mathrm{chem}})} \Delta t_{\mathrm{chem}} - n_{\mathrm{e}}^{(t_{\mathrm{chem}})} n_{\mathrm{ H\,II}}^{(t_{\mathrm{chem}})} \alpha_{\mathrm{ H}}(T) \Delta t_{\mathrm{chem}},
\end{equation}
where the photoionisation rate at $t_{\mathrm{chem}}$ is given by
\begin{equation}\label{phot_ion_rate_subcycle}
    \Gamma_{\mathrm{ H}}^{(t_{\mathrm{chem}})} = \Gamma_{\mathrm{ H}} \left( \frac{ 1-e^{- \tau^{(t_{\mathrm{chem}})}} }{ 1-e^{- \tau} } \right) \frac{ n_{\mathrm{ H\,I}} }{ n_{\mathrm{ H\,I}}^{(t_{\mathrm{chem}})} }, 
\end{equation}
where $\Gamma_{\mathrm{ H}}$ and $\tau$ are the photoionisation rate and optical depth at the beginning of the subcycling and $\tau^{(t_{\mathrm{chem}})} = \tau \, n_{\mathrm{ H\,I}}^{(t_{\mathrm{chem}})} / n_{\mathrm{ H\,I}} $. By defining the photionisation rate in this way, the ionising flux in the cell is constant during the radiative transfer time step. This subcycling scheme becomes computationally expensive when $\Delta t_{\mathrm{chem}} \ll \Delta t_{\mathrm{rt}}$, but photoionisation equilibrium is generally reached after a few subcycles. It is then no longer necessary to explicitly integrate the rate equation, but instead use the values of the preceding subcycle step.This way of subcycling ensures photon conservation even for large radiative transfer time steps.

\subsection{Time stepping}
As we discussed in Sect. \ref{section_transfer_equation} we are primarily interested in solving the time-independent radiative transfer equation, which means that the speed of light is assumed to be infinite. In all the tests presented in this paper, photons are moved only one Delaunay edge during a radiative transfer time step, thus interacting only at one grid point in that time step. We therefore have to be careful that the radiative transfer time step $\Delta t_{\mathrm{rt}}$ is sufficiently small to satisfy the time-independent transfer equation. In the limit that $\Delta t_{\mathrm{rt}}$ goes to zero, the time it takes for photons to leave the simulation box goes to zero as well, satisfying the condition of infinite speed of light. For all tests presented in this paper, we have checked that the radiative time step is sufficiently small to be in agreement with this limit.
\par The subcycling scheme that is used to calculate the evolution of the neutral fraction at a grid point during a radiative transfer time step allows for much larger time steps than needed to satisfy the time-independent transfer equation. This is very useful in simulations where the photons are allowed to travel more than one Delaunay edge per time step, for example in case one needs to solve the time-dependent transfer equation. However, this was not done for the tests presented in this paper.

%__________________________________________________________________

\section{Tests}\label{section_test_problems}
In order to test the accuracy of the new {\sc SimpleX} algorithm, we have performed several tests that were part of the Radiative Transfer Comparison Project \citep{Iliev:2006p13}. The original implementation of {\sc SimpleX} was part of this project but only did some of the tests. This gives us the opportunity to show the differences between the two versions in these tests, and show the behaviour of the new algorithm in a shadowing test that wasn't originally done by {\sc SimpleX} for the Comparison Project. For comparison purposes, in all tests presented here we adopt the on-the-spot approximation.
\par The increasing realism of the tests presented here allows us to highlight the different improvements of the method. The first test shows the importance of the direction conserving transport scheme to accurately account for the ionisations in the regions with low opacity and quantifies the optical depth at which it is necessary to switch from ballistic to direction conserving transport. The second test enables us to quantify the number of direction bins necessary to account for shadowing behind a dense cloud in direction conserving transport. Finally, in the third test we can assess the importance of the new sampling routine and study the minimum resolution required when dynamic grid updates are applied.

\subsection{Test 1: Isothermal \Hii region expansion}\label{section_test1}
One of the few problems in radiative transfer that has a known analytical solution is the \Hii region expansion in a homogeneous medium. A steady monochromatic source emits $\dot{N_{\gamma}}$ photons per second of frequency $h \nu = 13.6 \, \, \mbox{eV}$ into an initially neutral medium with constant gas density $n_{H}$. In equilibrium, the number of photons emitted by the source is balanced by the number of photons absorbed due to recombinations in a spherical volume. The radius at which equilibrium is reached is the Str\"{o}mgren radius, given by
\begin{equation}\label{stromgren_radius}
r_{\mathrm{S}} = \left( \frac{3 \dot{N_{\gamma}} }{ 4 \pi \alpha_{\mathrm{B}}(T) n_{\mathrm{H}}^{2} } \right)^{1/3}.
\end{equation}
Assuming an infinitely thin ionisation front and a fully ionised inner region, the ionisation front radius and velocity as function of time are
\begin{equation}\label{IFront_pos}
r_{I} = r_{\mathrm{S}} \left( 1 - e^{ -t/t_{\mathrm{rec} } } \right)^{1/3}
\end{equation}
and
\begin{equation}\label{IFront_vel}
v_{I} = \frac{ r_{\mathrm{S}} }{ 3 t_{\mathrm{rec} } } \frac{ e^{ -t/t_{\mathrm{rec} } } } { \left( 1 - e^{ -t/t_{\mathrm{rec} } }\right)^{2/3 } },
\end{equation}
shown as the black solid lines in Fig. \ref{figure_IFront_ballistic}.
\par We can improve on this by dropping the assumption of a fully ionised inner region of the Str\"{o}mgren sphere and calculating the neutral and ionised fraction as a function of radius by solving (e.g. \citet{Osterbrock:1989p2350} )
\begin{equation}\label{inner_stromgren_region}
\frac{ n_{\mathsc{H \, i}}}{4 \pi r^{2}} \int \mathrm{d} \nu \dot{N_{\gamma}}(\nu) e^{-\tau_{\nu}} \sigma_{\mathrm{H}}(\nu) = x^{2}(r)n_{\mathrm{H}}^{2}\alpha_{\mathrm{B}}(T).
\end{equation}
By using the commonly employed definition of the position of the ionisation front as the radius at which $x = 0.5$, solving this equation by direct integration gives us a second way of obtaining the Str\"{o}mgren radius. This yields a slightly different ionisation front position than obtained in Eq. (\ref{IFront_pos}). We show the solution obtained from directly integrating Eq. (\ref{inner_stromgren_region}) as the black dashed line in Fig. \ref{figure_IFront_ballistic}.
\par The analytical solutions thus obtained make this test ideally suited to test the different transport types of {\sc SimpleX} described in Sect.  \ref{section_radiation_transport}. We therefore performed this test problem with ballistic, direction conserving and combined transport to investigate the behaviour of the specific schemes. The parameters for this test are as follows. The computational box has length $ L = 13.2 \, \mathrm{kpc} $, the gas number density is $n_{\mathsc{H}} = 10^{-3} \, \mathrm{cm}^{-3}$, the temperature of the gas is $T = 10^{4} \, \mathrm{K}$. A source is placed in the centre of the box, emitting $\dot{N_{\gamma}} = 5 \cdot 10^{48} \, \mathrm{ionising} \,  \mathrm{photons} \,  \mathrm{s}^{-1} $. For these parameters, $t_{\mathrm{rec}} = 3.86 \cdot 10^{15} \, \mathrm{s} = 122.4 \, \mathrm{Myr} $ and $ r_{\mathrm{S} } = 5.4 \, \mathrm{kpc}$. The total simulation time is $500 \, \mathrm{Myr} \approx 4 t_{\mathrm{rec}}$. Note that this test differs slightly from Test 1 in \citet{Iliev:2006p13}, where the computational volume is smaller and the source is located in the corner of the computational box. Except where noted, a resolution of $64^{3}$ grid points and a time step of 0.05 Myr is used for this test. The grid on which we will perform this test is constructed by using the recipe described in Sect.  \ref{section_grid}, by using a homogeneous Poisson process to place the grid points. This introduces more shot noise compared to using a glass-like distribution, in which the point process is modified to make the points avoid one another. However, this is the same procedure we will apply for inhomogeneous density distributions using Eq. (\ref{realistic_sampling}), so in order to get a good understanding of the limitations of the method, we choose to use the Poisson process over a glass-like distribution for this test.

\subsubsection{Ballistic transport}\label{section_test1_sc}
The single mode of transport in the original {\sc SimpleX} algorithm was ballistic transport, described in Sect.  \ref{section_ballistic_transport}. This mode of transport was designed for regions where $\tau \ge 1$ between grid points. However, as the medium gets ionised during the simulation, the optical depth between grid points becomes so small that it is no longer correct to transport photons in this way. As described in KPCI09, a photon transported ballistically loses all memory of its initial direction after approximately 5 steps on the grid. Therefore, using ballistic transport in the highly ionised inner region of the Str\"{o}mgren sphere introduces numerical diffusion. 
\par The numerical diffusion does not influence the position of the ionisation front severely. 
\begin{figure}
   \centering
   \includegraphics[width=9cm]{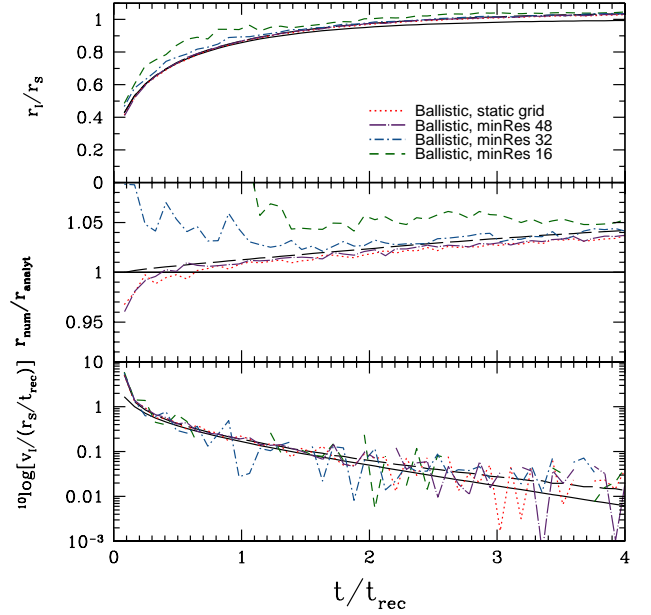}
      \caption{ The position, relative error and velocity of the ionisation front for Test 1. The black solid curves represent the analytic solutions of Eq. (\ref{IFront_pos}) and Eq. (\ref{IFront_vel}), while the black dashed curve represents the results of directly integrating Eq. (\ref{inner_stromgren_region}). Shown in colour are the results of {\sc SimpleX} simulations with ballistic transport only, where the red curves represent a simulation with a static grid and the violet, blue and green curves simulations with a dynamic grid with minimum resolutions of 48, 32 and 16, respectively. The position of the ionisation front is within 1\% of the analytical solution, although the effects of the limited resolution are clearly visible in the runs with a low minimum resolution.}
         \label{figure_IFront_ballistic}
\end{figure}
As is shown in Fig. \ref{figure_IFront_ballistic}, the red dotted line representing ballistic transport follows the numerical solution (the black dashed line) very closely, the error at the end of the simulation time is approximately 1 percent. The inner structure of the ionised region will be wrong, however, as we expect the numerical diffusion to dominate in the inner region of the Str\"{o}mgren sphere, were a large number of steps in an optically thin region needs to be taken, instead of close to the ionisation front. 
\begin{figure}
   \centering
   \includegraphics[width=8cm]{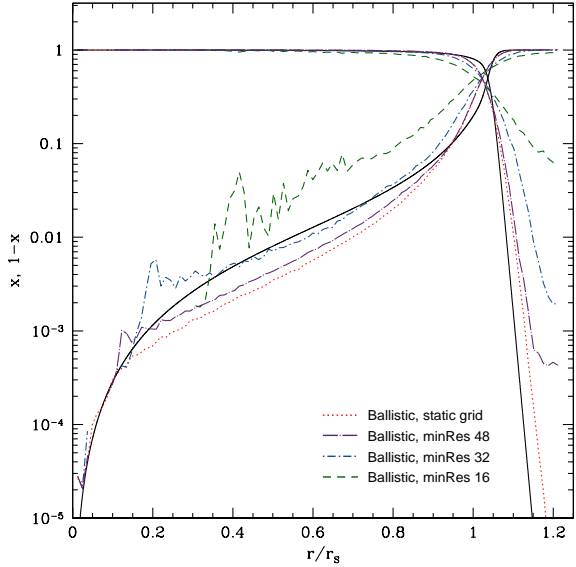}
      \caption{ Spherically averaged neutral and ionised fractions as function of the radial distance from the source after 500 Myr for Test 1. The black solid curve represents the result of directly integrating Eq. (\ref{inner_stromgren_region}). Shown in colour are the results of {\sc SimpleX} simulations with ballistic transport only, where the red curve represents a simulation with a static grid and the violet, blue and green curves represent simulations with a dynamic grid with minimum resolutions of 48, 32 and 16, respectively. Ballistic transport in the ionised regions results in numerical diffusion, therefore the neutral fraction is too low in the inner region, as can be seen from the dotted red curve. Removing grid points during the simulation, so essentially keeping the optical depth between grid points constant during the simulation, alleviates the diffusion problem, but introduces other errors as a result of the low resolution. The minimum resolution imposed gives a handle on how to control the errors stemming from numerical diffusion and a too low resolution.}
   \label{figure_IFractions_ballistic}
\end{figure}
In Fig. \ref{figure_IFractions_ballistic}, the spherically averaged neutral and ionised fractions are plotted as a function of distance from the source. The numerical diffusion in the inner region results in a too low neutral fraction. The reason for this is that the source photons quickly become diffuse and therefore, instead of travelling straight to the ionisation front, stay longer in the inner parts, thus cancelling more recombinations and causing a lower neutral fraction than expected from the analytical solution. 
\par One possible solution to this problem is removing grid points that have too low optical depths (see Sect.  \ref{section_grid_dynamics}). In Fig. \ref{figure_IFront_ballistic} and Fig. \ref{figure_IFractions_ballistic} the results are shown for simulations where grid points are removed until a certain minimum resolution is reached. Fig. \ref{figure_IFractions_ballistic} shows that the effect of numerical diffusion on the inner structure of the ionised region is lessened by the removal of grid points. The neutral fraction comes closer to the analytic solution as the minimum resolution decreases and the equilibrium position of the ionisation front becomes slightly more accurate. However, the slightly more accurate equilibrium results come at a cost. In Fig. \ref{figure_IFront_ballistic} we see that the lower resolution of 16 and 32 in the inner region causes the ionisation front position to deviate more than 5 percent from the analytic solution at early times, even though the equilibrium solution is accurate. Also, the spherically averaged equilibrium neutral fraction shows some severe artefacts due to the low resolution, most pronounced close to the source. Therefore, we conclude that only removing grid points with low optical depth is not a viable remedy against numerical diffusion, since the low resolution needed in the ionised regions causes severe noise in the equilibrium solution. 

\subsubsection{Direction conserving transport}
The numerical diffusion in ballistic transport is caused by the loss of direction of the photons after a number of interactions at grid points. In Sect.  \ref{section_direction_conserving_transport}, we described how we can cure this problem by defining solid angles in which the photons travel. The number of solid angles is a measure for the accuracy of the direction conservation. 
\begin{figure}
   \centering
   \includegraphics[width=8cm]{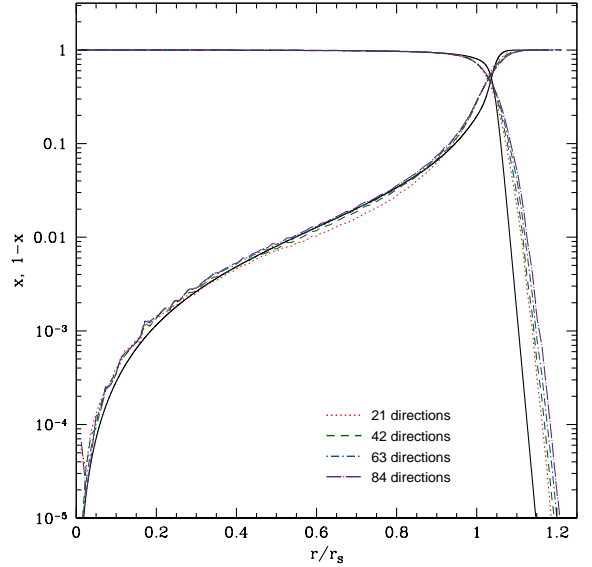}
      \caption{ Same layout as Fig. \ref{figure_IFractions_ballistic}. Shown in colour are the results of {\sc SimpleX} simulations with direction conserving transport, where the number of directions in which the photons are stored is 21, 42, 63 and 84. Clearly, the numerical diffusion of which ballistic transport suffers is absent if 42 directions or more are used to store the photons.}
   \label{figure_IFractions_LC}
\end{figure}
In Fig. \ref{figure_IFractions_LC}, the results of direction conserving transport with 21, 42, 63 and 84 direction bins are shown. From this plot we can see that the neutral fraction in the inner part of the Str\"{o}mgren sphere follows the analytical solution accurately if we use 42 direction bins or more. We can therefore conclude that the direction conserving transport scheme is an excellent solution for transporting photons in the optically thin regime. Since the difference between 42 direction bins and 63 and 84 bins is negligible, we are justified in using 42 direction bins when direction conserving transport is used in this test. In the next tests we will show the influence of the number of direction bins on the shadowing properties of the algorithm.
\par Fig. \ref{figure_IFractions_LC} also shows that with the use of fewer than 42 direction bins a small error from the numerical diffusion remains present in the simulation, the red dotted line representing the simulation using 21 direction bins deviates slightly from the analytic solution. The angular sampling does improve with the use of 21 direction bins compared to the average number of neighbours of a vertex (approximately equal to 15.54 in this point distribution), but the number of bins is too small to prevent photons from deviating from their original path.
\par The accurate transport of photons in the optically thin regime comes at a price. The direction conserving transport is computationally more expensive than ballistic transport, due to the fact that the direction bins need to be associated with every outgoing Delaunay line along which the photons are transported. To prevent preferential directions on the grid, the direction bins need to be rotated randomly after every time step, which causes additional computational overhead. Even though the transport of photons itself is almost as fast as with ballistic transport, the calculation of the grid properties takes more time. This extra computational cost can be reduced by combining both transport modes.

\subsubsection{Combined transport}\label{test1_CT_switchTau}
The introduction of direction bins prevents numerical diffusion in the optically thin regime but adds some computational overhead. Because the numerical diffusion is only present in the optically thin regime, we can speed up the calculation by combining ballistic and direction conserving transport in such a way that at high and moderate optical depth the faster ballistic transport is used, while vertices in the low optical regime employ the direction conserving mode of transport. This results in a computation time that is significantly faster than direction conserving transport in typical cosmological applications. 
\par The optical depth at which is switched from one mode of transport to the other is an important parameter. If it is too low, ballistic transport is done in regions with too low an optical depth, causing numerical diffusion. If it is too high, direction conserving transport is done in optically thick regimes, causing unnecessary computational overhead. 
\begin{figure}
   \centering
   \includegraphics[width=8cm]{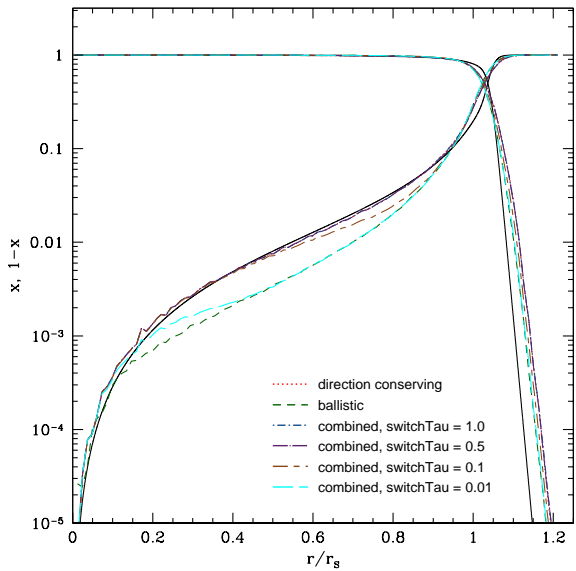}
      \caption{ Same as Fig. \ref{figure_IFractions_ballistic}. The coloured graphs represent the results of combined transport with different optical depts at which is switched from ballistic to direction conserving. If the switch is set too low, at $\tau =  0.01$ and $\tau = 0.1$, the numerical diffusion is not completely absent from the simulation. If we choose $\tau \ge 0.5$, the results are the same as with completely direction conserving transport.  }
   \label{figure_IFractions_CT_switchTau}
\end{figure}
In Fig. \ref{figure_IFractions_CT_switchTau}, the influence of the optical depth at which is switched is shown. If the conversion from ballistic to direction conserving transport happens at $\tau = 0.01$ and $\tau = 0.1$, we can see in this plot that there is still numerical diffusion in the inner region of the Str\"{o}mgren sphere. However, if the switch is made at $\tau \ge 0.5$, the difference between fully direction conserving and combined transport disappears. To be on the safe side, we use in the remaining tests a switch at $\tau = 1.0$. This way, we are sure that the numerical diffusion is completely absent in the simulations.
\par In order to get a good understanding of the behaviour of the new {\sc SimpleX} algorithm, we used the same time step of 0.05 Myr and the same resolution of $64^{3}$ grid points in all previous simulations. Using the combined transport scheme with the fiducial value of $\tau =  1.0$ for the switch between combined and direction conserving transport, we can proceed to find how large the influence of both the time step and the resolution is. 
\begin{figure}
   \centering
   \includegraphics[width=9cm]{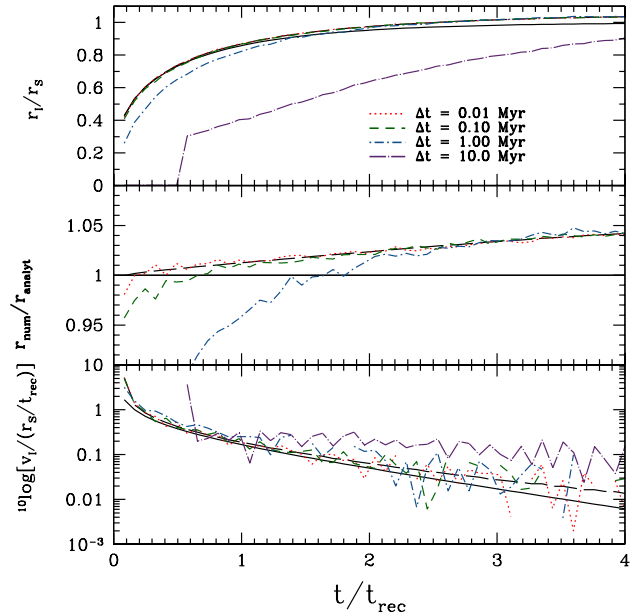}
      \caption{ Same as Fig. \ref{figure_IFront_ballistic}. The coloured graphs represent the different time steps of $\Delta t = 0.01, 0.1, 1 \, \mathrm{and} \, 10 \, \mathrm{Myr}$.}
   \label{figure_IFront_CT_time_step}
\end{figure}
In Fig. \ref{figure_IFront_CT_time_step} the position of the ionisation front as a function of time is plotted for different simulation time steps $\Delta t$. Since photons can travel only from one grid point to another during a time step, large time steps show unphysical behaviour, especially when the equilibrium solution has not been reached yet. From Fig. \ref{figure_IFront_CT_time_step} we see that at a time step of $\Delta t = 1 \, \mathrm{Myr}$ the ionisation front is behind the analytical solution at early times, but it gives the correct equilibrium solution. Time steps smaller than that retrieve the analytical solution to within 1\%. A time step of 10 Myr gives an ionisation front position that is behind even at equilibrium, because for these large time steps photons are travelling slower than the speed of light. Another reason is that in order to prevent preferential directions on the grid, the direction bins need to be randomly rotated every time step. If the time step is too large, there are not enough rotations to avoid these preferential directions, resulting in an incorrect equilibrium solution. This issue could be solved by letting photons travel more than one Delaunay edge in a time step. However, in that case the only difference between the simulations would be the time at which the physics is be evaluated, which is not what we were interested in for this comparison. 
\par
\begin{figure}
   \centering
   \includegraphics[width=8cm]{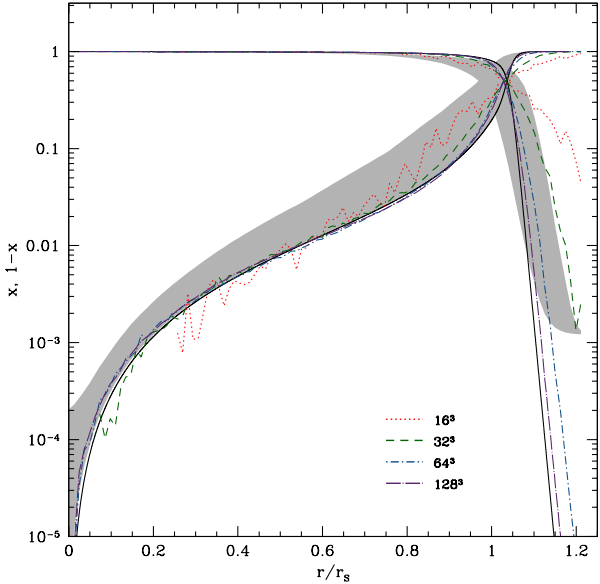}
      \caption{ Same as Fig. \ref{figure_IFractions_ballistic}. The coloured graphs represent simulations with a different resolution of $16^{3}, 32^{3}, 64^{3} \, \mathrm{and} \, 128^{3}$. The shaded area represents the range of neutral and ionised fractions found by the codes in the Radiative Transfer Comparison Project \citep{Iliev:2006p13}. All {\sc SimpleX} simulations lie within the shaded area, showing that even at very low resolution {\sc SimpleX} gives acceptable results.} 
   \label{figure_IFractions_CT_resolution}
\end{figure}

 Using a time step of 0.05 Myr, Fig. \ref{figure_IFractions_CT_resolution} shows the effect of the resolution on the neutral and ionised fraction as function of radius and compares this to the results of other codes in the Radiative Transfer Comparison Project, shown as the shaded area.
For all simulations, the ionisation front is at the same position. For low resolution, the neutral fraction in the inner part of the Str\"{o}mgren sphere is noisy, but still lies within the range of solutions reported by other codes. Higher resolution simulations are less noisy and converge to the same solution. 

\subsection{Test 2: Shadowing behind a dense cloud}\label{section_shadowing}
The second test in the Radiative Transfer Comparison Project examines ionisation front trapping in a dense clump and the formation of a shadow. Unfortunately, this test involves a plane-parallel wave-front, which is something that is difficult to impose in {\sc SimpleX}. The reason for this is that a plane-parallel wave as described in this test represents a preferential direction, which is exactly what we try to avoid in our method. It is possible to alter the method to produce a plane-parallel wave. However, it would then be unclear how much this alteration affects the shadowing properties in other applications where this alteration is not used. 
\par We therefore chose to do a different shadowing test, similar to the one described in \citet{Mellema:2006p22}. This test involves the irradiation of a dense clump by a point source. The test set-up is similar to the test described in the previous section, except that we put a dense, uniform, rectangular slab with size 855 by  2138 by 2138 kpc at a distance of 1096 kpc from the source. The density contrast between the homogenous environment and the clump is $n_{\mathrm{clump}}/n_{\mathrm{out}} = 200$. The ionisation front will be trapped inside the dense clump and a shadow should form behind the clump. 
\par
\begin{figure}
   \centering
   \includegraphics[width=8cm]{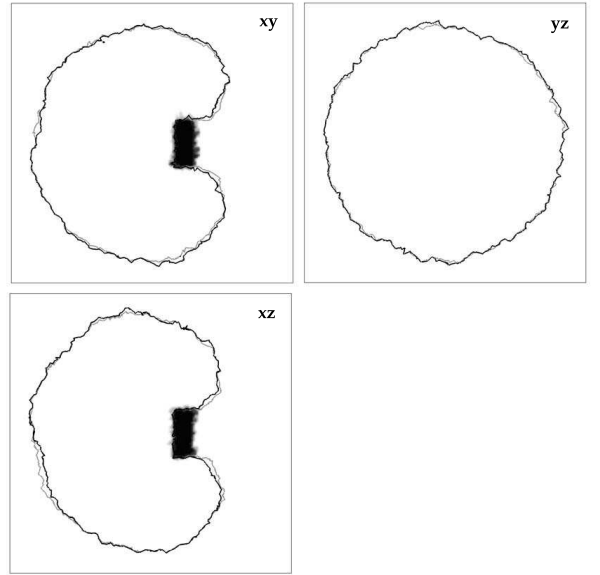}
      \caption{Test 2 results for a resolution of $64^{3}$ grid points. Shown are slices along the xy-plane, xz-plane and yz-plane, as indicated. The dense clump is shown in black. The contours represent the position of the ionisation front at the end of the simulation, $t = 500 \, \mathrm{Myr}$. The grey contours represent the simulation where the photons are stored in 42 direction bins at optically thin vertices, while the black contours represent the simulation where 84 direction bins were used. At this resolution, the increased angular resolution of the photons in the simulation with 84 direction bins does not have a significant impact on the shadowing. The ionisation front penetrates approximately 5 cells into the region shadowed by the clump. The slice through the x-axis shows that the ionisation front outside the shadow is not affected by the dense clump.}
      \label{figure_test2_64}
\end{figure}

 This test provides an excellent means to study the number of direction bins in which the photons have to be stored at optically thin vertices to ensure proper shadowing. In Fig. \ref{figure_test2_64} the results of this test at a resolution of $64^{3}$ grid points is shown.
At this resolution the effect of different angular samplings is negligible, the shadow that is cast has the same width for both the simulation involving 42 direction bins and 84 direction bins. The ionisation front penetrates approximately 5 cells into the region shadowed by the clump. This is a result of the fact that photons are transported along the $d$ most straightforward Delaunay edges of the triangulation. Some of these edges are pointing into the region shadowed by the clump, thereby ionising the medium inside the shadowed region. By restricting the angle in which to look for the $d$ most straightforward neighbours (which is $90^{\circ}$ in these simulations, see Sect.  \ref{section_radiation_transport}), the sharpness of the shadow will gradually increase. However, this leads to an ionised region that is no longer spherical in the regions that are not shadowed, because the source no longer 'fills' the entire sky with radiation. 
\par
\begin{figure}
   \centering
   \includegraphics[width=8cm]{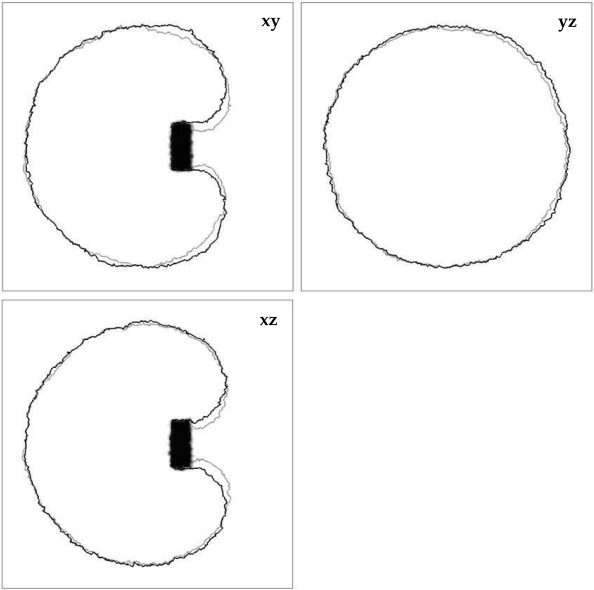}
      \caption{Same as Fig. \ref {figure_test2_64} but with a resolution of  $128^{3}$ grid points. At this resolution, the increased angular resolution of the photons in the simulation with 84 direction bins does have a significant impact on the shadowing. With 84 direction bins, the ionisation front penetrates approximately 8 cells into the region shadowed by the clump, while with 42 direction bins this is almost twice as many cells. The slice through the x-axis shows that the ionisation front outside the shadow is not affected by the dense clump.}
      \label{figure_test2_128}
\end{figure}

 Fig. \ref{figure_test2_128} shows the same test at a resolution of $128^{3}$ grid points. 
At this resolution the effect of using 84 direction bins instead of 42 is profound. While with 84 direction bins photons penetrate approximately 8 cells into the shadowed region, with 42 direction bins this is approximately 15 cells. The shadow cast in case of 84 direction bins is comparable to the shadow in the simulation using $64^{3}$ cells, the width being governed by the $d$ most straightforward directions. This is not the case when 42 direction bins are used to store the photons at optically thin vertices. There is a small amount of diffusion associated with every rotation of the direction bins. At a resolution of $128^{3}$ grid points, the photons undergo so many rotations on their path that this diffusion starts to be visible in the shadowing properties of the method. This unwanted numerical scattering is cured by using more direction bins. 
\par This shadowing test has shown that the direction conserving transport scheme at optically thin vertices provides the means to cast shadows behind dense clumps. As a result of the choice to send photons along the edges of the Delaunay triangulation, {\sc SimpleX} will never be able to reproduce the infinitely sharp shadows that ray tracing methods produce. The random rotation of the direction bins at every radiative transfer time step that is necessary to avoid preferential directions on the grid causes a small amount of numerical diffusion that shows up when photons traverse many optically thin vertices. We have shown that by increasing the number of direction bins this problem is cured. This test shows the importance of removing redundant optically thin vertices in highly ionised regions, as described in Sect.  \ref{section_grid_dynamics}, to minimise the number of optically thin vertices that need to be traversed.

\subsection{Test 3: Multiple sources in a cosmological density field}\label{section_test_4}
The final test we conducted is closest to our intended application, that of a cosmological density field. It is for this kind of problem that the {\sc SimpleX} method has been primarily designed, since the test consists of multiple sources in a density field with a large dynamic range. The initial conditions are given by a time slice at $z=9$ from a cosmological N-body and gas-dynamic simulation using the cosmological PM+TVD code \citep{Ryu:1993p2454}. The box size is $0.5 h^{-1} \mathrm{Mpc}$, the gas temperature is initially set to 100 K. The sources  are located in the 16 most massive haloes in the box, emitting $f_{\gamma} = 250 $ ionising photons  per atom over $t_{s} = 3 \, \mathrm{Myr}$ resulting in a photon flux of 
\begin{equation}
  \dot{N}_{\gamma} = f_{\gamma} \frac{M \Omega_{b}}{\Omega_{0} m_{H} t_{s}},
\end{equation} 
where $M$ is the total halo mass, $\Omega_{0} = 0.27$, $\Omega_{b} = 0.043$ and $h = 0.7$. The total simulation time is 0.4 Myr. {\sc SimpleX} does not solve for the temperature state of the gas, so instead we assume a temperature of $1 \cdot 10^{4} \, \mathrm{K}$ for the ionised gas. This might cause differences in the number of recombinations compared to the other codes in the comparison project that do solve for the temperature, since the recombination rate is a function of the temperature.

\subsubsection{Sampling function} 
In order to perform this test, we first have to translate from the grid-based representation of the density field to the {\sc SimpleX} grid. As discussed in Sect.  \ref{section_realistic_sampling}, it is essential to have the highest dynamic range possible while keeping the density gradients to a minimum. Referring to Eq. (40) in KPCI09, we set the sampling parameter $Q_{n}$ to 5. We choose the parameter $\alpha$ in Eq. (\ref{realistic_sampling}) close to its maximal value of 0.3 in order to have the highest resolution possible for this $Q_{n}$ in the low density regions. This results in $n_{0} = 3.69 \cdot 10^{-5} \, cm^{-3}$. The point density in the lowest density regions is equivalent to a resolution of approximately $77^{3}$ for $128^{3}$ grid points. 
\par
\begin{figure*}
   \centering
   \includegraphics[width=\textwidth]{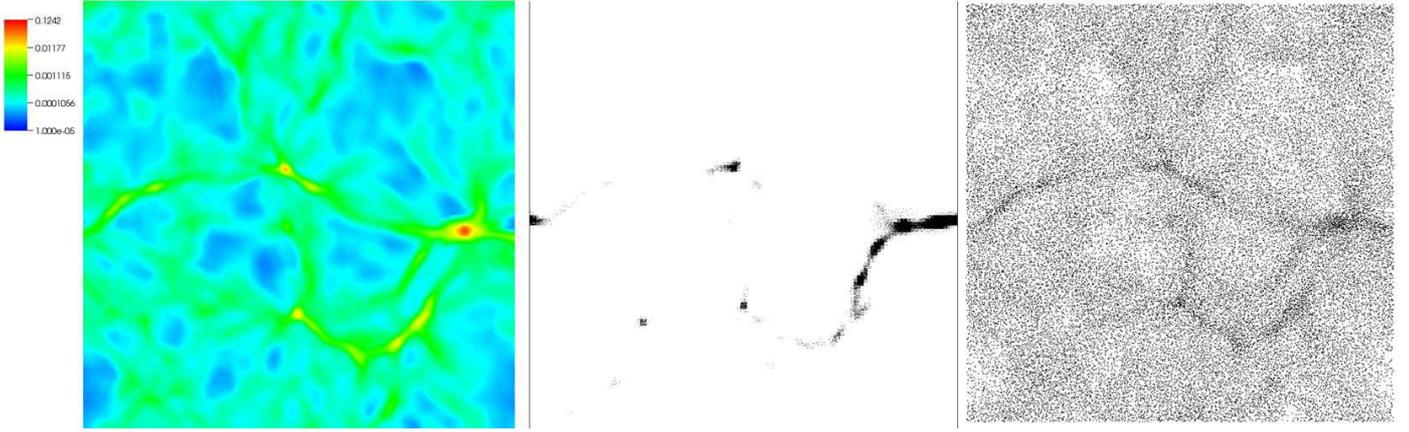}
      \caption{\textit{Left:} Slice through the cosmological density field of \citet{Iliev:2006p13} at coordinate $z = z_{\mathrm{box}}/2$. \textit{Centre:} Cubic sampling using $128^{3}$ grid points, as required by Eq. (\ref{delaunay_vs_mfp}), showing that for realistic density fluctuations the low and intermediate density regions are severely undersampled. \textit{Right:} Sampling scheme described in Sect.  \ref{section_realistic_sampling}, with parameters $\alpha = 0.3$ and  $\rho_{0} = 3.69 \cdot 10^{-5}$. The point density in the lowest density regions corresponds approximately to a resolution of $77^{3}$.}
   \label{figure_density_sampling}
\end{figure*}

 A slice through the $z = z_{\mathrm{box}}/2$ coordinate of the grid with the above parameters is shown on the right hand side in Fig. \ref{figure_density_sampling}.
If we compare the hybrid sampling scheme to the cubic sampling scheme, we can clearly see that the new scheme produces a grid that has the desired higher resolution in the dense filaments, but still has enough grid points in the low density regions to ensure that photons can travel into these regions. It is this grid that we will use for performing the radiative transfer simulations.

\subsubsection{The result of undersampling}\label{section_test4_undersampling}
\begin{figure}
   \centering
   \includegraphics[width=9cm]{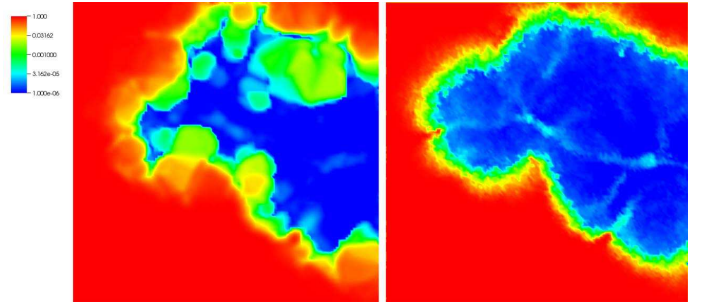}
      \caption{Comparison between the result with the old sampling routine as was performed for the Radiative Transfer Comparison Project (left) and the hybrid sampling scheme (right). The number of grid points is in both cases $128^{3}$, the mode of transport is ballistic only. For comparison reasons both grids have been interpolated to a regular grid of $128^{3}$ cells. Shown is a slice through the $z = z_{\mathrm{box}}/2$ coordinate of the computational domain at $t = 0.2$ Myr, as the influence of the incorrect sampling is most pronounce there. In the left hand plot, the artefacts of undersampling the low density regions are visible as neutral clumps, where no radiation has gone yet. The reason for this is the small number of grid points in these regions, resulting in a too small number of photons travelling there. The plot on the right hand side shows that this problem is solved by correctly sampling the low density regions.}
   \label{figure_old_vs_new}
\end{figure}

To stress the importance of sampling the medium correctly, we show in Fig. \ref{figure_old_vs_new} the results using the old sampling method, used by {\sc SimpleX} for the Radiative Transfer Comparison Project, and the hybrid sampling.
For comparison purposes the mode of transport in both cases is ballistic and the number of grid points is $128^{3}$. The result of the incorrect sampling is clearly visible as dense neutral clumps in the ionised regions. This is not due to the fact that photons have preferential directions into the dense filaments, otherwise we would also see this effect in the hybrid sampling case. Rather, it is caused by the fact that there are too few grid points in the low density regions, resulting in very large cells. Since photons travel along the Delaunay edges, radiation simply does not travel into the low density regions, resulting in the observed large neutral clumps in the voids. 
\par
\begin{figure}
   \centering
   \includegraphics[width=9cm]{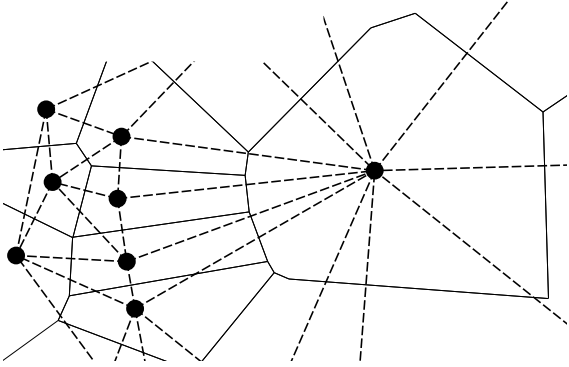}
      \caption{Example of the effect of large grid cells on radiation transport. Consider radiation travelling from left to right along the edges of the triangulation. Even though there are many Delaunay edges pointing towards the big cell, this cell will not be ionised easily. The reason for this is that radiation is sent to the $d$ most straightforward neighbours and the photons have no memory of their original direction. In this case approximately half of the photons travelling from right to left that should be used to ionise the big cell, will instead be travelling around this cell. Weighting the most straightforward direction would alleviate this problem, as would restricting the opening angle in which the $d$ most straightforward neighbours are allowed to be.}
   \label{figure_big_cell}
\end{figure}

 An example of this effect is shown in Fig. \ref{figure_big_cell}.
Consider photons travelling from left to right along the edges of the triangulation. The reason why the large cell is hard to ionise is not that there are not enough Delaunay edges pointing towards the cell. Clearly, the number of edges pointing towards the big cell is above average. However, as photons are sent to the $d$ most straightforward neighbours and have no memory of their original direction, approximately half of the photons will be travelling around the big cell instead of ionising it. Thus, large cells are harder to ionise, resulting in the neutral clumps in the low density regions visible on the left-hand side of Fig. \ref{figure_old_vs_new}. This problem would be partly cured by the use of weights of the $d$ most straightforward neighbours, giving the edges pointing into the big cell a higher number of photons. Another option would be to restrict the opening angle in which the $d$ most straightforward neighbours are allowed to be, thus discarding most of the edges that point around the big cell in case of photons travelling from the right. Applying the direction conserving transport scheme will not solve this problem entirely, since the photons are still split up and travelling along the edges of the triangulation as with ballistic transport, so the problem is essentially the same. However, a slightly larger fraction of photons will be travelling into the big cell with direction conserving transport compared to ballistic transport, as photons that are send in a direction around the large cell remember their original direction and thus have a higher probability to travel into the direction of the big cell again.

\subsubsection{Grid dynamics}\label{section_test4_dynamics}
\begin{figure}
   \centering
   \includegraphics[width=9cm]{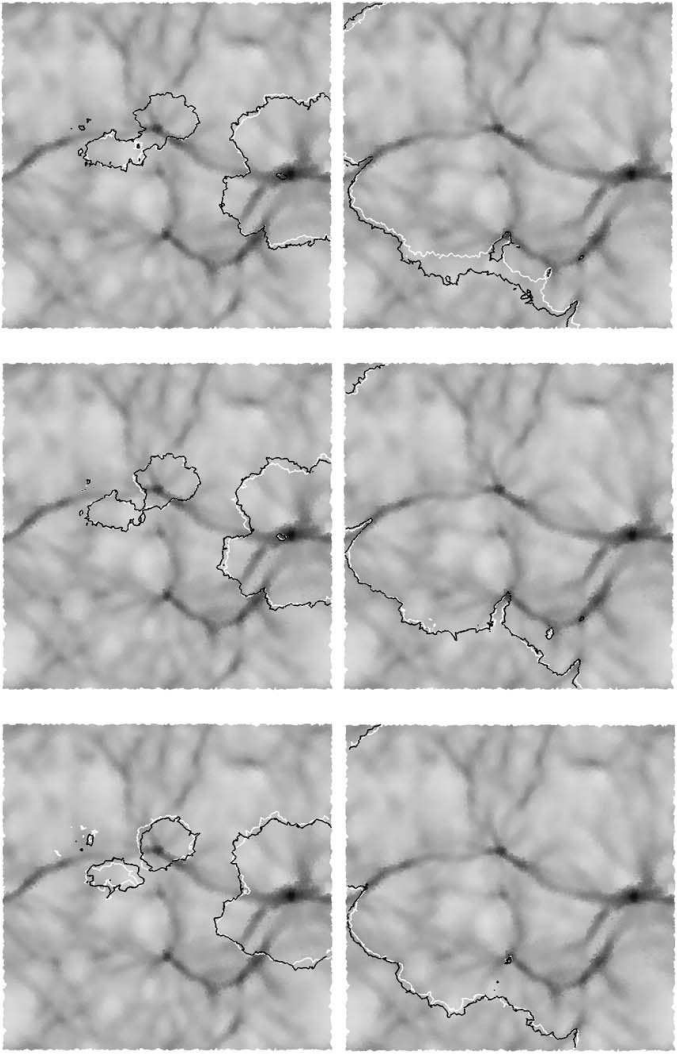}
      \caption{ Test 3, slice through coordinate $z = z_{\mathrm{box}}/2$ of the cosmological density field. Contours show the position of the ionisation front for simulations using 42 direction bins at optically thin vertices (white) and 84 direction bins (black). Shown are the results of simulations without vertex removal (top), vertex removal with a minimum local resolution of $128^{3}$ (centre) and vertex removal with a minimum local resolution of $77^{3}$ (bottom). The left panels show the ionisation front after 0.05 Myr, the right panels after 0.4 Myr. The removal of vertices in highly ionised regions reduces the numerical scatter due to the random rotations of the direction bins, therefore the difference between the two simulations in the plot in the centre stays within a few percent. The centre plot also shows sharper shadows than the plot on the left-hand side. The right-hand side plot shows that when the minimum resolution is very low, the medium is not well represented anymore, causing the shadows to disappear.}
   \label{figure_test4_128_updates}
\end{figure}

We have performed a set of simulations to investigate the effect of regularly updating the grid according the changes in the optical depth. In Fig. \ref{figure_test4_128_updates} six different simulations are shown. 
We performed one simulation where no vertices were removed, one simulation where highly ionised vertices were removed until a minimum local resolution of $128^{3}$ was reached and one simulation where the minimum resolution of the vertex removal was $77^{3}$. Thus, the second simulation removes redundant optically thin vertices until the resolution of the input hydrodynamics grid has been reached, removing the additional resolution of the adaptive grid when it's no longer necessary, while the third simulation removes redundant optically thin vertices until the minimum resolution of the {\sc SimpleX} grid itself has been reached. In all three cases we also varied the number of direction bins in which the photons at optically thin vertices were stored between 42 and 84 bins. 
\par When we compare the lower panels of Fig. \ref{figure_test4_128_updates} which show the position of the ionisation front at the end of the simulation time, a clear difference is visible between the simulations. The centre plot shows that the removal of redundant grid points until the resolution of the original hydrodynamics grid has been reached results in sharper shadows at the place of high density filaments, because the small amount of scattering due to the random rotations of the direction bins has been kept to a minimum. Moreover, in this simulation the difference between using 42 and 84 direction bins is only a few percent, showing that the numerical scatter is indeed negligible. The effect of vertex removal is not apparent at earlier times in the simulation (figure \ref{figure_test4_128_updates}, top panel). In this case, the ionised regions occupy a small volume in the computational box so the photons do not have to travel a large distance. Only at a later stage, when photons have to travel almost an entire box length, the influence of the numerical scatter is visible. The effect of the scatter remains small even at these late stages of the simulation, only visible near high density filaments. 
\par In the simulations shown on the right-hand side of Fig. \ref{figure_test4_128_updates} is visible that the minimum resolution should not be lower than the resolution of the original input grid. In that case we are not only removing the extra grid points that were put in the high density filaments due to the adaptive nature of the {\sc SimpleX} grid, we are also removing grid points that are necessary to represent the physical structures in the original hydrodynamics grid. By removing too many grid points from the high density filaments, these structures are smeared out over a few large cells in the {\sc SimpleX} grid. Hence, recombinations occurring in the filaments are spread out over too large a volume, resulting in a lack of shadowing in the dense filaments.

\subsubsection{Comparison to other codes}
\begin{figure}
   \centering
   \includegraphics[width=9cm]{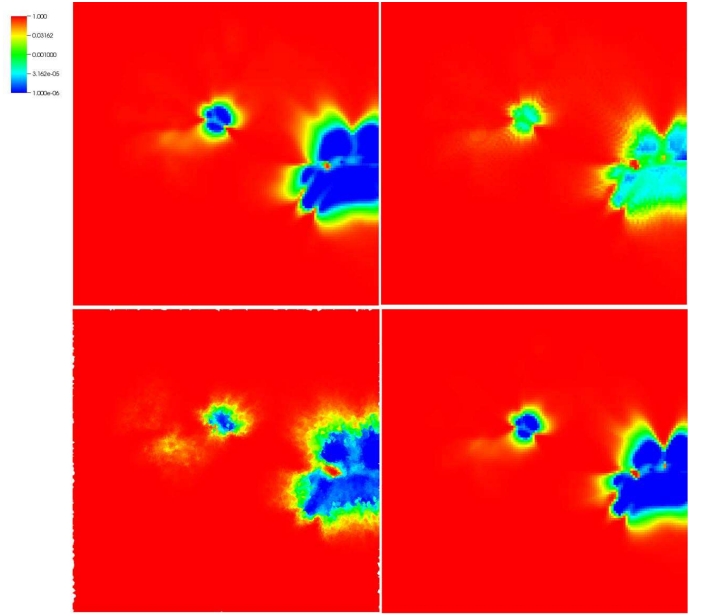}
      \caption{Slice through the $z = z_{\mathrm{box}}/2$ coordinate of the cosmological density field showing the \Hi fraction at time $t = 0.05 \, \mathrm{Myr}$. Beginning in the top left hand corner and proceeding clock-wise are the results from $\mathsc{c}^{2}\mathsc{ray}$, {\sc crash}, {\sc ftte} and {\sc SimpleX}.}
   \label{figure_test4_compare_05}
\end{figure}

\begin{figure}
   \centering
   \includegraphics[width=9cm]{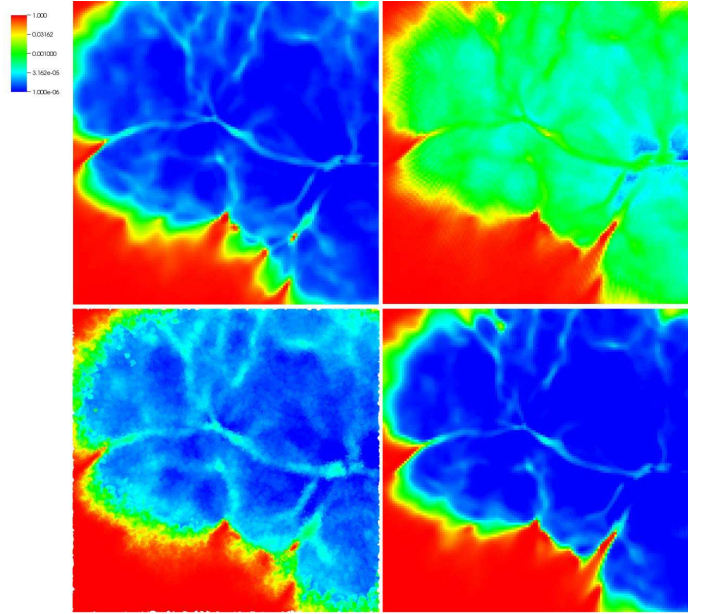}
      \caption{ Slice through the $z = z_{\mathrm{box}}/2$ coordinate of the cosmological density field showing the \Hi fraction at time $t = 0.4 \, \mathrm{Myr}$, the final simulation time. Beginning in the top left hand corner and proceeding clock-wise are the results from $\mathsc{c}^{2}\mathsc{ray}$, {\sc crash}, {\sc ftte} and {\sc SimpleX}. Note that with the direction conserving transport, {\sc SimpleX} is capable of producing sharp shadows behind dense structures.}
   \label{figure_test4_compare_40}
\end{figure}

The {\sc SimpleX} run with vertex removal and a minimum local resolution of $128^{3}$ is compared to the results of other codes in the Radiative Transfer Comparison Project in Fig. \ref{figure_test4_compare_05} and Fig. \ref{figure_test4_compare_40}.   
Shown is the \Hi fraction of a slice through the $z = z_{\mathrm{box}}/2$ coordinate of the computational volume at times $t=0.05$ Myr and $t = 0.4$ Myr. The results of the {\sc SimpleX} simulation are compared to those of $\mathsc{c}^{2}\mathsc{ray}$ \citep{Mellema:2006p22}, {\sc crash} \citep{Maselli:2003p983} and {\sc ftte} \citep{Razoumov:2005p2463}. Note that the new version of the {\sc crash} code, described in \citet{Maselli:2009p990}, shows results that are in better agreement with the $\mathsc{c}^{2}\mathsc{ray}$ and {\sc ftte} results. 
\par Both the position of the ionisation front as the inner structure of the ionised regions of the {\sc SimpleX} simulation show very good agreement with the other codes, despite the fact that {\sc SimpleX} does not solve for the temperature state of the gas. In the {\sc SimpleX} simulation the temperature of the ionised gas is constant at $10^{4}$ K. This is slightly lower than the temperature that the other codes, which do solve for the temperature, find. Thus, {\sc SimpleX} overestimates the number of recombinations that are occurring in the ionised gas. This results in an ionisation front that is slightly behind that of the other codes at the end of the simulation. Also visible in these figures is the effect of Poisson noise on the stochastic grid, causing a less smooth ionisation front. This effect decreases with higher resolution. We are currently investigating how to minimise the Poisson noise during the creation of the grid. 
\par With the direction conserving transport in the ionised regions, {\sc SimpleX} is capable of reproducing the shadows behind neutral clumps and filaments that the other codes show as well. These shadows were absent in the simulation with ballistic transport only, the right hand side of Fig. \ref{figure_old_vs_new}. On the whole, the morphology of the ionised region that {\sc SimpleX} finds shows excellent agreement with the results of the other codes.

%%______________________________________________________________

\section{Summary} \label{section_conclusions}
In this paper we have presented essential updates to the original {\sc SimpleX} algorithm described in \citet{Ritzerveld:2006p9}. The most important improvement are the parallellisation for distributed memory machines, the new sampling function and the transport mode for optically thin regions. 
\par All the radiation transport algorithms used in {\sc SimpleX} are local, photons travel from one grid point to the other. This makes the method relatively straightforward to parallellise. We have shown that the parallel execution of {\sc SimpleX} shows excellent scaling properties on distributed memory machines, the work load per processor is decreased by almost 50\% when the number of processors on which the algorithm is executed is doubled. An exception to this fortunate scaling is the triangulation algorithm, due to the extra points in the boundary around processors that need to be triangulated. For the present application this is not a problem, since the computation time of the triangulation is an order of magnitude smaller than the computation time of the radiative transfer itself.
\par The new sampling functions allows us to effectively choose the gradient in point density as to get an idea of the numerical diffusion that might occur. Together with a choice for the minimum resolution in the low density regions, this results in a grid optimal for doing radiative transfer calculations with {\sc SimpleX}. We have shown that by constructing the grid in the correct way, the artefacts that showed up with the original {\sc SimpleX} algorithm when doing simulations of a realistic density field with large density fluctuations are absent. 
\par The direction conserving transport mode ensures that photons keep their directions when they are travelling through regions with a low optical depth. This removes the numerical diffusion that occurs when photons are transported ballistically everywhere. Furthermore, with direction conserving transport {\sc SimpleX} results show sharp shadows behind dense regions. Even though direction conserving transport is computationally more expensive than ballistic transport, by applying it only to the regions where it is necessary, the extra computational overhead is limited. 
\par We would like to stress that these improvements in the accuracy of the method have been achieved without sacrificing its main benefits. The algorithm remains computationally very efficient, with a computation time independent of the number of sources. Due to the adaptive nature of the grid, the dynamic range of the resolution is very high. Combined with the parallellisation for distributed memory machines, simulations can be done with a very high numbers of grid points, making {\sc SimpleX} an ideal tool for doing large scale reionisation simulations.
\par We have shown that {\sc SimpleX} produces very accurate results in various standard test problems for cosmological reionisation. In the classical case of a growing \Hii region about a single source both the position of the ionisation front and the inner ionised region agree with the analytic expectations within 1\%. In the test problem closest to our future application, that of a cosmological density field with multiple sources, the {\sc SimpleX} results show an excellent agreement with the results of other codes.

\begin{acknowledgements}
The authors would like to thank Milan Raicevic, Tom Theuns, Tilo Beyer and Michael Meyer-Hermann for useful discussions and Jakob van Bethlehem and Rien van de Weygaert for carefully reading the manuscript. JPP thanks ICC Durham and the University of Frankfurt for their hospitality. JPP acknowledges support from the European Science Foundation (ESF) for the activity entitled 'Computational Astrophysics and Cosmology'.
\end{acknowledgements}


\begin{thebibliography}{}

\bibitem[{Abel {et~al.}(1999)Abel, Norman, \& Madau}]{Abel:1999p992}
Abel, T., Norman, M.~L., \& Madau, P. 1999, The Astrophysical Journal, 523, 66

\bibitem[{Barber {et~al.}(1995)Barber, Dobkin, \& Huhdanpaa}]{Barber:1995p2617}
Barber, C.~B., Dobkin, D.~P., \& Huhdanpaa, H. 1995, ACM Transactions on
  Mathematical Software, 22, 469

\bibitem[{Cen(2002)}]{Cen:2002p2314}
Cen, R. 2002, The Astrophysical Journal Supplement Series, 141, 211

\bibitem[{Delaunay(1934)}]{Delaunay:1934p2614}
Delaunay, B. 1934, Classe des Sciences Mathematiques et Naturelles, 7, 793

\bibitem[{Dirichlet(1850)}]{Dirichlet:1850p2615}
Dirichlet, G.~L. 1850, Journal f{\"u}r die reine und angewandte Mathematik, 40,
  209

\bibitem[{Gnedin \& Abel(2001)}]{Gnedin:2001p24}
Gnedin, N.~Y. \& Abel, T. 2001, New Astronomy, 6, 437

\bibitem[{Gnedin \& Ostriker(1997)}]{Gnedin:1997p2439}
Gnedin, N.~Y. \& Ostriker, J.~P. 1997, Astrophysical Journal v.486, 486, 581

\bibitem[{Iliev {et~al.}(2006)Iliev, Ciardi, Alvarez, Maselli, Ferrara, Gnedin,
  Mellema, Nakamoto, Norman, Razoumov, Rijkhorst, Ritzerveld, Shapiro, Susa,
  Umemura, \& Whalen}]{Iliev:2006p13}
Iliev, I.~T., Ciardi, B., Alvarez, M.~A., {et~al.} 2006, MNRAS,
  371, 1057

\bibitem[{Kruip {et~al.}(2009)}]{Kruip:2009}
Kruip, C.~J.~H., Paardekooper, J.-P., Clauwens, B., Icke, V. 2009, A\&A submitted

\bibitem[{Kunasz \& Auer(1988)}]{Kunasz:1988p2312}
Kunasz, P. \& Auer, L.~H. 1988, J. Quant. Spectrosc. Radiat. Transfer, 39, 67

\bibitem[{Maselli {et~al.}(2009)Maselli, Ciardi, \& Kanekar}]{Maselli:2009p990}
Maselli, A., Ciardi, B., \& Kanekar, A. 2009, MNRAS, 393, 171

\bibitem[{Maselli {et~al.}(2003)Maselli, Ferrara, \& Ciardi}]{Maselli:2003p983}
Maselli, A., Ferrara, A., \& Ciardi, B. 2003, MNRAS, 345, 379

\bibitem[{Mellema {et~al.}(2006)Mellema, Iliev, Alvarez, \&
  Shapiro}]{Mellema:2006p22}
Mellema, G., Iliev, I., Alvarez, M., \& Shapiro, P. 2006, New Astronomy, 11,
  374

\bibitem[{Mihalas \& Mihalas(1984)}]{Mihalas:1984p2300}
Mihalas, D. \& Mihalas, B.~W. 1984, Foundations of Radiation Hydrodynamics, Oxford University Press, New York

\bibitem[{Miles(1970)}]{Miles1970}
Miles, R. E. 1970, Izvestilia Akademii Nauk Armlianskoi SSR Matematika, 5, 263

\bibitem[{Miles(1974)}]{Miles1974}
Miles, R. E. 1974, A synopsis of 'Poisson flats in Euclidian space', Stochastic Geometry (New York: John Wiley), 202-227 

\bibitem[{M{\o}ller(1989)}]{Moller1989}
M{\o}ller, J. 1989, Advances in Applied Probability, 21, 37

\bibitem[{Norman {et~al.}(1998)Norman, Paschos, \& Abel}]{Norman:1998p2308}
Norman, M.~L., Paschos, P., \& Abel, T. 1998, Memorie della Societa Astronomia
  Italiana, 69, 455

\bibitem[{Okabe(2000)}]{Okabe:2000p2282}
Okabe, A., Boots, B., Sigihara, K., \& Chiu, S. N. 2000, Spatial tessellations : concepts and applications of voronoi
  diagrams. 2nd ed. Chichester, John Wiley {\&} Sons Ltd

\bibitem[{Osterbrock \& Ferland(2006)}]{Osterbrock:1989p2350}
Osterbrock, D.~E., \& Ferland, G. J. 2006, Astrophysics of Gaseous Nebulae and Active Galactic Nuclei, 
  2nd edn. Univ. Science Books, Sausolito, CA

\bibitem[{Pawlik \& Schaye(2008)}]{Pawlik:2008p888}
Pawlik, A.~H. \& Schaye, J. 2008, MNRAS, 389, 651

\bibitem[{Razoumov \& Cardall(2005)}]{Razoumov:2005p2463}
Razoumov, A.~O. \& Cardall, C.~Y. 2005, MNRAS, 362, 1413

\bibitem[{Rijkhorst {et~al.}(2006)Rijkhorst, Plewa, Dubey, \&
  Mellema}]{Rijkhorst:2006p896}
Rijkhorst, E.-J., Plewa, T., Dubey, A., \& Mellema, G. 2006, A{\&}A, 452, 907

\bibitem[{Ritzerveld(2005)}]{Ritzerveld:2005p17}
Ritzerveld, J. 2005, A{\&}A, 439, L23

\bibitem[{Ritzerveld \& Icke(2006)}]{Ritzerveld:2006p9}
Ritzerveld, J. \& Icke, V. 2006, Physical Review E, 74, 26704

\bibitem[{Ritzerveld(2007)}]{Ritzerveld:2007p1304}
Ritzerveld, N. G. H. 2007, PhD thesis, Leiden Observatory, Leiden University,
P.O. Box 9513, 2300 RA Leiden, The Netherlands

\bibitem[{Ryu {et~al.}(1993)Ryu, Ostriker, Kang, \& Cen}]{Ryu:1993p2454}
Ryu, D., Ostriker, J.~P., Kang, H., \& Cen, R. 1993, Astrophysical Journal,
  414, 1

\bibitem[{Shirokov \& Bertschinger(2005)}]{Shirokov:2005p2410}
Shirokov, A. \& Bertschinger, E. 2005, arXiv, astro-ph:0505087

\bibitem[{Springel(2005)}]{Springel:2005p2411}
Springel, V. 2005, MNRAS, 364, 1105

\bibitem[{Springel(2010)}]{Springel:2010p1514}
Springel, V. 2010, MNRAS, 401, 791

\bibitem[{Trac \& Cen(2007)}]{Trac:2007p1516}
Trac, H. \& Cen, R. 2007, The Astrophysical Journal, 671, 1

\bibitem[{Voronoi(1908)}]{Voronoi:1908p2612}
Voronoi, G. 1908, Journal f{\"u}r die reine und angewandte Mathematik, 134, 198
      
\end{thebibliography}
\end{document}